\documentclass[referee]{aa}
\usepackage{graphicx}
\usepackage{psfig}
\usepackage{epsfig}
\usepackage{txfonts}
\begin{document}
\title{High resolution spectroscopy of the hot post-AGB stars :
IRAS13266-5551 (CPD-55 5588) and IRAS17311-4924 (Hen3-1428)
\thanks{Based on observations made with the Victor M. Blanco 4m
telescope of the Cerro Tololo Inter-American Observatory,
Chile.}}
\author{G. Sarkar \inst{1} \thanks{Present address : Indian Institute of Technology,
Kanpur - 208016, India}
\and M. Parthasarathy \inst{2}
\and Bacham E. Reddy \inst{2} \thanks{Visiting Observer, 
Cerro Tololo Inter-American Observatory, which is operated
by the Association of Universities for Research in Astronomy
Inc., under contract with the U.S. National Science Foundation.}}
\institute{Inter-University Centre for Astronomy and Astrophysics,
Post Bag 4, Ganeshkhind,\\ 
Pune - 411007, India
\and Indian Institute of Astrophysics, Koramangala, Bangalore - 560034,
India}
\offprints{G. Sarkar
\email{gsarkar@iitk.ac.in}}
\date{Received / Accepted}
\authorrunning{G. Sarkar et al.}
\titlerunning{Hot post-AGB stars : IRAS13266-5551 and IRAS17311-4924}

\abstract{The high resolution spectra covering the wavelength range
4900 \AA~ to 8250 \AA~ of the hot post-AGB stars
IRAS13266-5551 (CPD-55 5588) and IRAS17311-4924 (Hen3-1428) 
reveal absorption lines of C~II, N~II, O~II, Al~III, Si~III and Fe~III
and a rich emission line spectrum consisting of H~I, He~I, C~II, N~I, O~I, 
Mg~II, Al~II, Si~II, V~I, Mn~I, Fe~III, [Fe~II] and [Cr~II].
The presence of [N~II] and [O~I] lines and absence of [O~III] indicate low 
excitation nebulae around these stars. The components of Na~I absorption lines 
indicate the presence of neutral circumstellar envelopes in addition to
the low excitation nebulae around these two hot post-AGB stars. The H$_\alpha$ lines 
show P-Cygni profiles indicating ongoing post-AGB mass loss. 
From the absorption lines, we derived heliocentric radial velocities of
65.31 $\pm$ 0.34 km s$^{-1}$ and 27.55 $\pm$ 0.74 km s$^{-1}$ for 
IRAS13266-5551 and IRAS17311-4924 respectively. 
High galactic latitude and large radial velocity 
of IRAS13266-5551 indicate that it belongs to 
the old disk population. Preliminary 
estimates for the CNO abundances in IRAS13266-5551 are
obtained. 
}
\maketitle

\section{Introduction}
From the study of IRAS sources with far-IR colours similar to
that of planetary nebulae (PNe) several cool and hot post-AGB
stars have been discovered (Parthasarathy and Pottasch, 1986;
Parthasarathy et al., 2000a, Parthasarathy et al., 2001) forming an 
evolutionary sequence in the transition region from the tip of the AGB to the early
stages of PNe (Parthasarathy, 1993a, b). IRAS13266-5551 (CPD-55 5588) and 
IRAS17311-4924 (Hen3-1428) were identified as hot post-AGB stars (Table 1) 
based on their far-IR flux distribution, high galactic latitudes and 
B-supergiant spectra in the optical (Parthasarathy \& Pottasch, 1989; 
Parthasarathy, 1993a; Parthasarathy et al., 2000a). The UV (IUE) spectra of these
stars show C~II (1335 \AA~), Si~IV (1394 \AA~, 1403 \AA~), C~IV (1550 \AA~) 
and N~IV (1718 \AA~) lines typical of the central stars of PNe.
The C~IV (1550 \AA~) resonance lines are blue shifted 
indicating stellar wind velocities of $-$1821 km s$^{-1}$ (CPD-55 5588) 
and $-$1066 km s$^{-1}$ (Hen3-1428) respectively (Gauba and Parthasarathy,
2003). The ``30 $\mu$ feature", SiC emission at 11.5 $\mu$, 
and UIR band at 7.7 $\mu$ were detected in the ISO spectrum of 
IRAS17311-4924 (Gauba and Parthasarathy, 2004). 
These features have been detected in the circumstellar dust shells of 
carbon rich AGB stars (C-stars), post-AGB stars, proto-planetary
nebulae (PPNe) and planetary nebulae (PNe) (see e.g. Hony et al., 2002; 
Hrivnak et al., 2000, Volk et al., 2000, 2002). Loup et al. (1990)
detected CO emission in IRAS17311-4924 typical for
circumstellar shells around evolved objects. 

High resolution optical spectra of only a few hot post-AGB stars
have been analysed. These include IRAS01005+7910 (Klochkova et al., 2002),
IRAS18062+2410 (SAO85766, Parthasarathy et al., 2000b; Arkhipova et al., 2001a;  
Mooney et al., 2002, Ryans et al., 2003), IRAS19590-1249 (LSIV$-$12$^{\circ}$111, 
McCausland et al., 1992; Conlon et al., 1993a; Ryans et al., 2003) and IRAS20462+3416 
(LSII$+$34$^{\circ}$26, Parthasarathy, 1993b; Garc\'ia-Lario et al., 1997;
Arkhipova et al., 2001b). 
The optical spectra of these stars show absorption lines due to C~II, N~II, O~II, 
Si~II, Si~III, Fe~III etc. Emission lines of He~I, Fe~I, II and III, 
N~I, Ni~I, O~I have also been detected. Nebular emission lines of 
[O~II], [N~II], [S~II] etc., detached cold circumstellar dust shells, 
OB-supergiant spectral types, high galactic latitudes and chemical
composition indicate that these are PPNe
(Parthasarathy et al., 1993c, 1995 and 2000b). 
IRAS13266-5551 and IRAS17311-4924 are
found to be similar to the objects mentioned above. In this paper
we report an analysis of their high resolution spectra.

\begin{table}
\begin{center}
\caption{Details of the stars}
\begin{tabular}{c|c|c|c|c|c|c|c|c|c|c|c|c}
\hline
IRAS & Name & RA & DEC & l & b & Sp.Type & V & B-V &
\multicolumn{4}{c}{IRAS Fluxes (Jy.)} \\
     &      &  (2000) & (2000) &  & & Optical & & & 12 $\mu$ & 25 $\mu$ & 60 $\mu$ & 100 $\mu$ \\ 
\hline \hline
13266-5551 & CPD-55 5588 & 13:29:50.8 & -56:06:53 & 308.30 & +6.36 & B1Ibe & 10.68$^{a}$ &
0.31$^{a}$ & 0.76 & 35.90 & 35.43 & 11.66 \\ \hline
17311-4924 & Hen3-1428 & 17:35:02.49 & -49:26:26.4 & 341.41 & -9.04 & B1IIe & 10.68$^{b}$ &
0.40$^{b}$ & 18.34 & 150.70 & 58.74 & 17.78 \\ \hline

\end{tabular}

\noindent \parbox{14cm}{Photometry is from : $^{a}$Reed (1998) and $^{b}$Kozok (1985).\\
Spectral types are from Parthasarathy et al. (2000a).}
\end{center}
\end{table}           

\section{Observations}

High resolution (R $\sim$ 30,000) spectra of IRAS13266-5551 
and IRAS17311-4924 from 4900 \AA~ to 8250 \AA~ were obtained
on 22nd June, 2002. Each object was observed twice during the night. 
The echelle spectrograph at the f/7.8 Ritchey-Chretien focus of the 
Victor M. Blanco 4m. telescope of the Cerro Tololo Inter-American
Observatory (CTIO), Chile was used for the purpose. The spectra were recorded
using a Tektronix 2048X2048 CCD. The slit width was 150 $\mu$ corresponding to 
1\arcsec~ on the sky. Appropriate number of bias frames
and flat fields were observed. A Th-Ar comparison lamp was used for
wavelength calibration.

As these are hot stars, spectra in the blue would contain more 
number of absorption lines. However, our observing program and
the spectrograph setup did not allow us to go shortward of
4900 \AA~. Therefore, the analysis reported in this paper is
based on the spectra covering the wavelength range 4900 \AA~
to 8250 \AA~.

\section{Analysis}

The spectra were processed using standard IRAF routines.
They were corrected using data in the overscan region of the CCD
chip. The other reduction steps included trimming, bias subtraction,
flat field correction, correction for scattered light and wavelength
calibration. The two sets of reduced spectra for each object were then
combined to increase the signal-to-noise (S/N) ratio. The final S/N
ratios for IRAS13266-5551 and IRAS17311-4924 
were estimated to be $\sim$ 120. The reduced spectra were continuum
normalised and the equivalent widths (W$_{\rm \lambda}$) of the absorption
and emission lines were measured. Whenever required, deblending 
was done to obtain gaussian fits to the blended line profiles. The
continuum normalised spectra are presented in Appendices A and B
(Figs. A and B). The line identifications (Tables 2a, 2b, 2c, 2d, 3a, 3b, 3c and 3d) 
are based on the Moore multiplet table (1945) and the linelists of Parthasarathy
et al. (2000b) and Klochkova et al. (2002). Unidentified 
lines are denoted by ``UN". Night sky emission lines (atmospheric emission) were
identified from Osterbrock and Martel (1992) and Osterbrock et al. (1996) and are
listed as ``atmos." in the tables. The laboratory wavelengths, log (gf) values and 
excitation potentials ($\chi$) are from the linelist compiled by
Ivan Hubeny and retrieved from the directory /pub/hubeny/synplot  by
anonymous ftp from tlusty.gsfc.nasa.gov.

\subsection{Description of the spectra}

The high resolution optical spectra of IRAS13266-5551 and
IRAS17311-4924 show absorption lines due to C~II, N~II, O~II, 
Ne~I, Al~III and Si~III. The O~I triplet at $\sim$ 7773 \AA~ was detected
in both stars. Both stars show a rich emission line
spectrum with lines of C~II, Mg~II, Al~II, Si~II, 
Fe~III and [Cr~II] in emission. Emission lines of 
N~I, O~I, [O~I], V~I, Mn~I and [Fe~II] in IRAS17311-4924  
were also detected. The presence of low excitation nebular lines of 
[N~II] in the spectra of both stars and the absence of 
[O~III] 5007 \AA~ indicate that photoionisation has just started.

The He~I lines in the two stars show a variety of profiles. They 
appear in absorption, in emission and also show P-Cygni 
profiles indicating post-AGB mass-loss. The He~I(4) 5015.678 \AA~ 
and He~I(45) 7281.349 \AA~ emission lines in IRAS13266-5551   
are superposed on the corresponding absorption components. The 
asymmetric nature of these emission lines suggests that they may have
P-Cygni profiles. The presence of high excitation lines of He~I 
and low excitation emission lines of Na~I (see Sec. 3.4) and 
V~I indicate a range of temperatures for the 
circumstellar material around these stars. The circumstellar envelope 
around these stars may be extended and the outermost regions may be
cooler. The H$_{\alpha}$ lines in both stars show P-Cygni profiles. 
The emission peak of the line in IRAS17311-4924 
is asymmetric. 

\subsection{Radial velocities}

Heliocentric radial velocities (V$_{\rm r}$) for the well defined absorption 
and emission lines are presented in Tables 2a, 2b, 2d, 3a, 3b and 3d. 
The radial velocities of the Fe~III (5) absorption lines in IRAS13266-5551 are 
relatively larger than the rest and those of the Ne~I absorption lines in IRAS17311-4924 
are relatively smaller suggesting that these lines may be formed in different 
regions in the atmospheres of these stars.
Therefore, in estimating the mean heliocentric radial velocities, we have
excluded the above lines. The radial velocity of C~II (2) 6578.052 \AA~
absorption line has also been neglected (see footnote to Table 2a). 
We obtained mean radial velocities of 65.31 $\pm$ 0.34 km s$^{-1}$  
and 27.55 $\pm$ 0.74 km s$^{-1}$ from the absorption lines in IRAS13266-5551
and IRAS17311-4924 respectively. The mean heliocentric radial velocities of the emission 
lines are 58.32 $\pm$ 0.65 km s$^{-1}$ and 32.74 $\pm$ 0.43 for IRAS13266-5551 
and IRAS17311-4924 respectively. In estimating 
the mean radial velocity of the emission lines we have excluded the radial velocity 
measurements of the forbidden lines. The errors given here refer to the probable errors
of estimation. Figs. 1 a and b show the overall radial 
velocity trend for the absorption and emission lines with respect to the 
equivalent widths (W$_{\rm \lambda}$) and lower excitation potentials of these lines. 

The mean heliocentric radial velocity from absorption lines in the
case of IRAS17311-4924 corresponds to V$_{\rm LSR}$ = 31.13 km s$^{-1}$. This
value may be compared with the velocity (V$_{\rm LSR}$ =) of 
36 km s$^{-1}$ derived from CO observations of the star by Loup et al. (1990). 

\setcounter{figure}{0}
\begin{figure}
\renewcommand{\thefigure}{\arabic{figure}a.}
\epsfig{figure=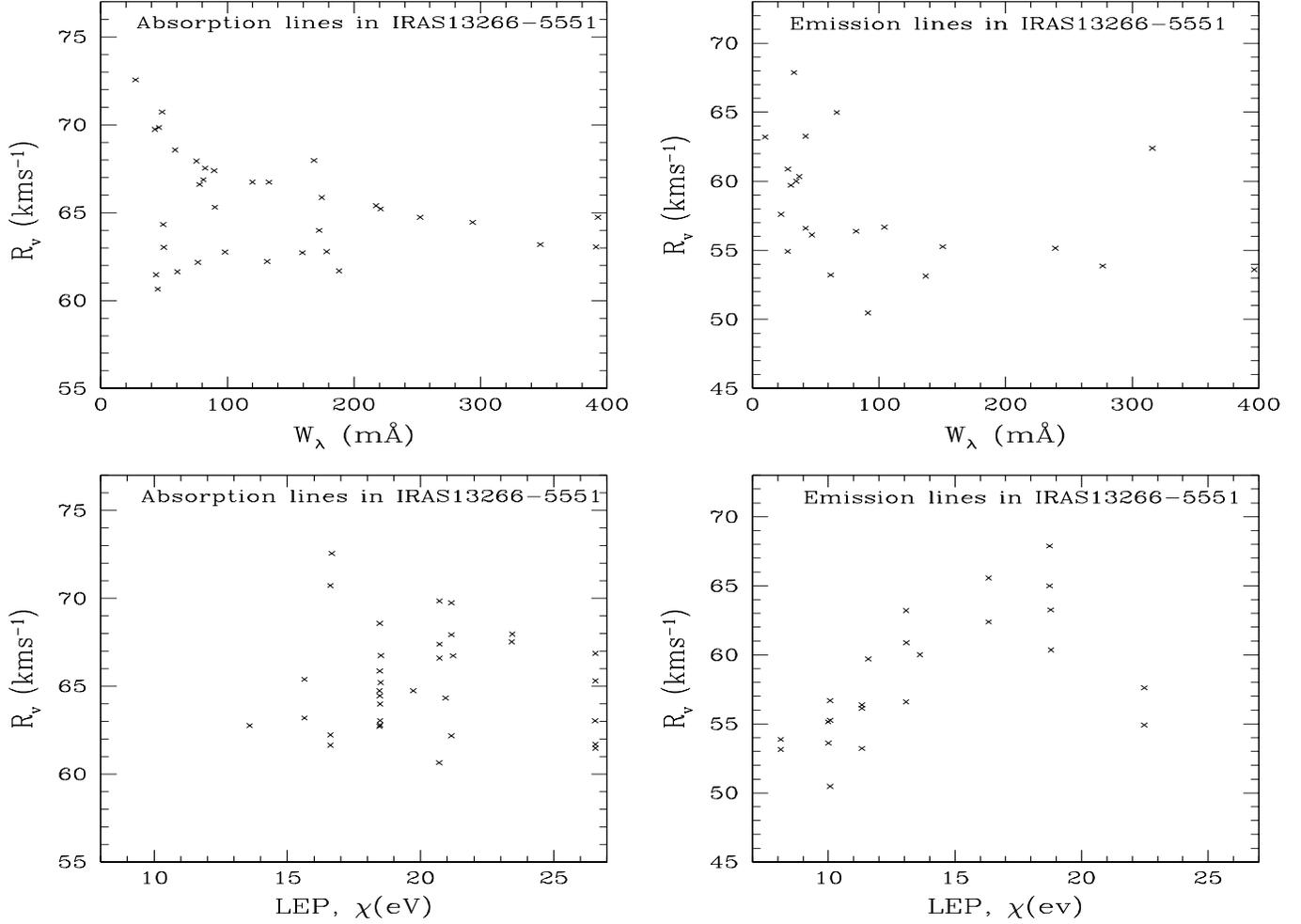, width=18cm, height=13cm}
\caption{Radial velocity trends of the absorption and emission 
lines in IRAS13266-5551. Radial velocity measurements of the forbidden
lines have not been plotted.}
\end{figure}

\setcounter{figure}{0}
\begin{figure}
\renewcommand{\thefigure}{\arabic{figure}b.}
\epsfig{figure=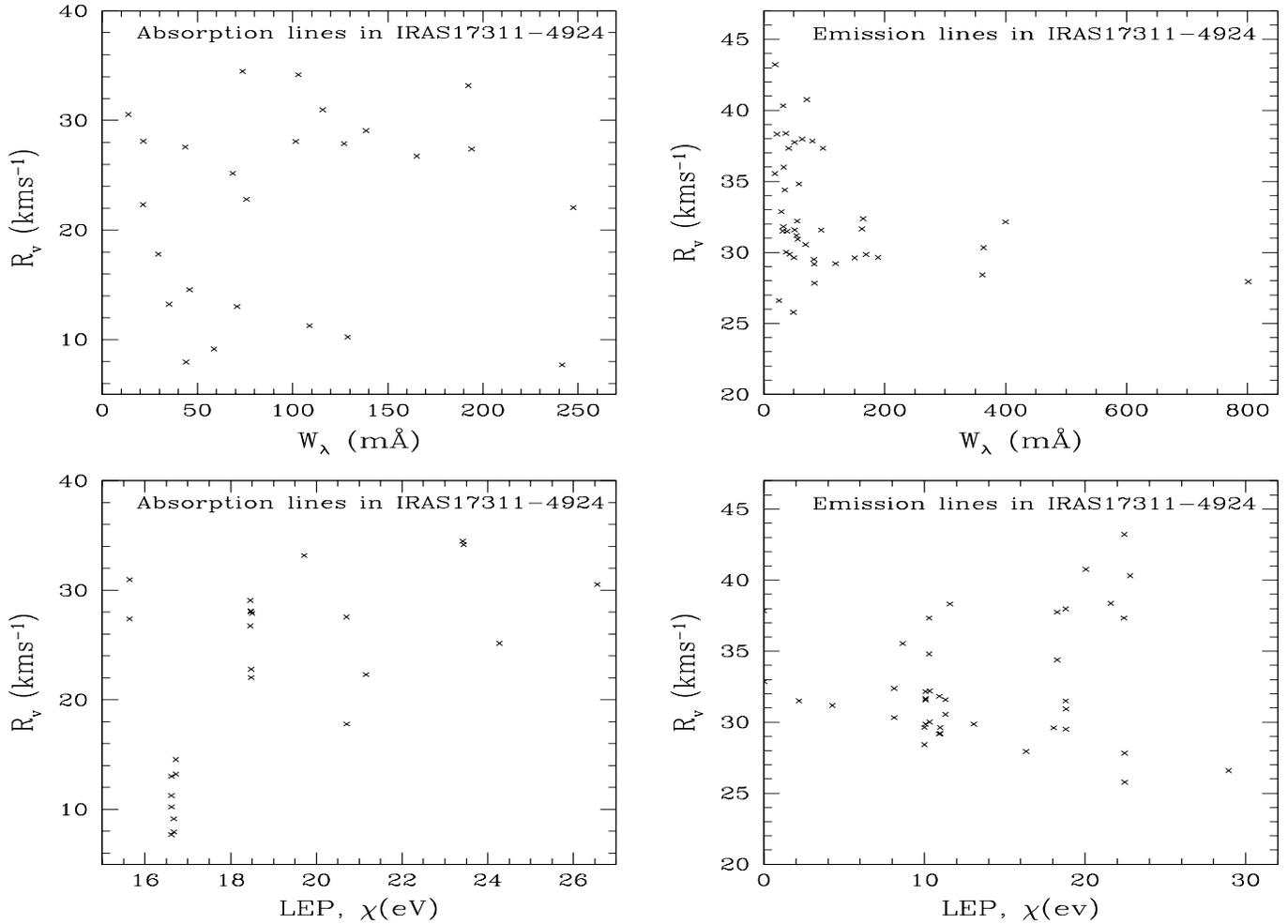, width=18cm, height=13cm}
\caption{Radial velocity trends of the absorption and emission lines 
in IRAS17311-4924. Radial velocity measurements of the forbidden lines
have not been plotted.}
\end{figure}

\subsection{Wind velocities from the P-Cygni profiles}

We estimated wind velocities from the well defined and
unblended blue absorption edges of the P-Cygni profiles 
of He~I, C~II and Fe~III (Tables 2c and 3c). The 
absorption components of the H$_{\alpha}$ P-Cygni profiles
are affected by the broad wings of the H$_{\alpha}$ emission
components (Fig. 3) and hence could not be used to estimate the
wind velocities in these stars. The absorption
component of He~I(45) 7281.349 \AA~ P-Cygni profile in
IRAS17311-4924 may be affected by atmospheric absorption lines 
in this region. The wind velocities in IRAS17311-4924 
increase with the LEP of the species involved. 

\subsection{Diffuse interstellar bands (DIBs)}

DIBs are absorption features in the spectra of reddened 
stars and have their origin in the interstellar and 
circumstellar medium. They are typically broader than expected 
from the Doppler broadening of turbulent gas motions in the interstellar and
circumstellar medium. DIB at 5780.410 \AA~ was identified in the spectra of 
IRAS13266-5551 and IRAS17311-4924. IRAS13266-5551
also exhibited DIBs at ${\rm \lambda \rm \lambda}$ 5797.030 \AA~, 6195.990 \AA~,
6203.060 \AA~ and 6613.630 \AA~. Their heliocentric radial velocities
(V$_{\rm r}$, Tables 2a and 3a) rule out the possibility of circumstellar
origin. From the strength of the band at 5780.410 \AA~ (W$_{\rm \lambda}$ = 169.6 m\AA~ 
for IRAS 13266-5551 and 109.9 m\AA~ for IRAS17311-4924),
we estimate interstellar E(B$-$V) $\simeq$ 0.35 and 0.20 respectively
(Herbig, 1993). At the galactic latitude and longitude of these stars, 
using the Diffuse Infrared Background Experiment (DIRBE)/IRAS dust maps 
(Schlegel et al., 1998), we estimated interstellar extinction values of 
0.53 and 0.22 respectively. Herbig (1993) concludes  that although the DIB 
strengths increase linearly with E(B$-$V), there is a real dispersion about 
the mean relationship. For example from their data, HD144470 and HD37061 both have
about the same DIB strengths despite large differences in their
values of E(B$-$V), 0.22 and 0.56 respectively.

\subsection{Na~I D$_{2}$ and Na~I D$_{1}$ lines}

The Na~I D$_{2}$ and Na~I D$_{1}$ lines in the spectra of 
IRAS13266-5551 and IRAS17311-4924  
show both absorption and emission components (Fig. 2, Tables 2d and 3d). 
Comparing the radial velocities of these components with the average
radial velocities for each star (Sec. 3.2), it is evident that the 
absorption components 1,2 and 3 in IRAS13266-5551 and 
component 1 in IRAS17311-4924 are of interstellar origin. 
Components 4 and 2 in IRAS13266-5551 and IRAS17311-4924 respectively
may be of circumstellar origin indicating the
presence of neutral circumstellar envelopes in addition to
the cold detached dust shells and low excitation nebulae. 

Radial velocities of the emission components of the Na~I lines are 
very close to the mean radial velocity of emission lines in 
IRAS13266-5551 (= $+$58.32 km s$^{-1}$) and 
IRAS17311-4924 (= $+$32.74  km s$^{-1}$). However, the emission
components of the Na~I lines may be disturbed by closely 
located absorption components and therefore the velocities 
derived from the Na~I emissions may be in error.

\begin{figure}
\epsfig{figure=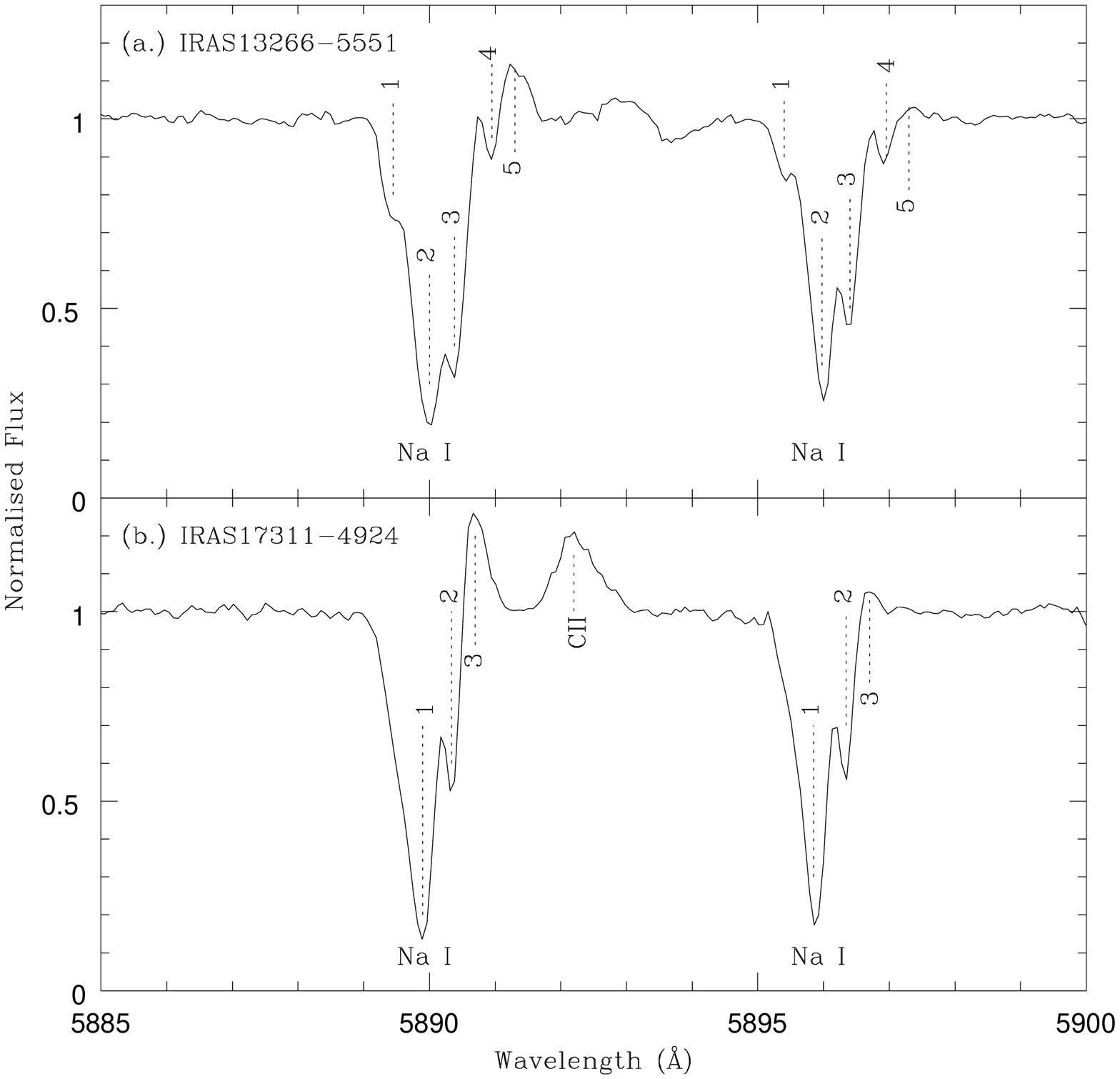}
\caption{Na~I D$_{2}$ and Na~I D$_{1}$ lines in the spectra of 
(a.) IRAS13266-5551 and (b.) IRAS17311-4924. 
The various absorption and emission components
of the lines have been labelled.}
\end{figure}

\subsection{H$_{\alpha}$ profile and mass loss rate}

H$_{\alpha}$ profiles of the two stars are shown in Fig. 3. 
The wavelengths were converted to velocity units relative to
the laboratory wavelength of the H$_{\alpha}$ line, 6562.817 \AA~ .
The zero point was then adjusted for the heliocentric radial
velocity of each star. The wings of H$_{\alpha}$
can be seen upto about 220 and 180 km s$^{-1}$ in IRAS13266-5551 
and IRAS17311-4924 respectively.

The P-Cygni profile of H$_{\alpha}$ indicates ongoing mass-loss.
Model calculations by Klein and Castor (1978) predicted a
tight relationship between the H$_{\alpha}$ luminosity
of the stellar wind and mass loss rate. The H$_{\alpha}$ 
luminosity is related to the equivalent width of the H$_{\alpha}$
emission line (see e.g. Conti and Frost, 1977; Ebbets, 1982). 
The equivalent widths of the H$_{\alpha}$ emission components
are 6.902 \AA~ and 9.872 \AA~ in IRAS13266-5551 
and IRAS17311-4924 respectively (Tables 2c and 3c).
Modelling the H$_{\alpha}$ profiles to derive the mass loss
rates of post-AGB stars would be the subject of a future paper.
Here, we may compare our H$_{\alpha}$ profiles with the B1.5Ia star, 
BD-14$^{\circ}$ 5037. The observed equivalent width of
the H$_{\alpha}$ emission component in this star is 7.4 \AA~.  
From the H$_{\alpha}$ profile, Leitherer (1988) derived
a mass loss rate of 1.58X10$^{-6}$ M$_{\odot}$ yr$^{-1}$. 

\begin{figure}
\epsfig{figure=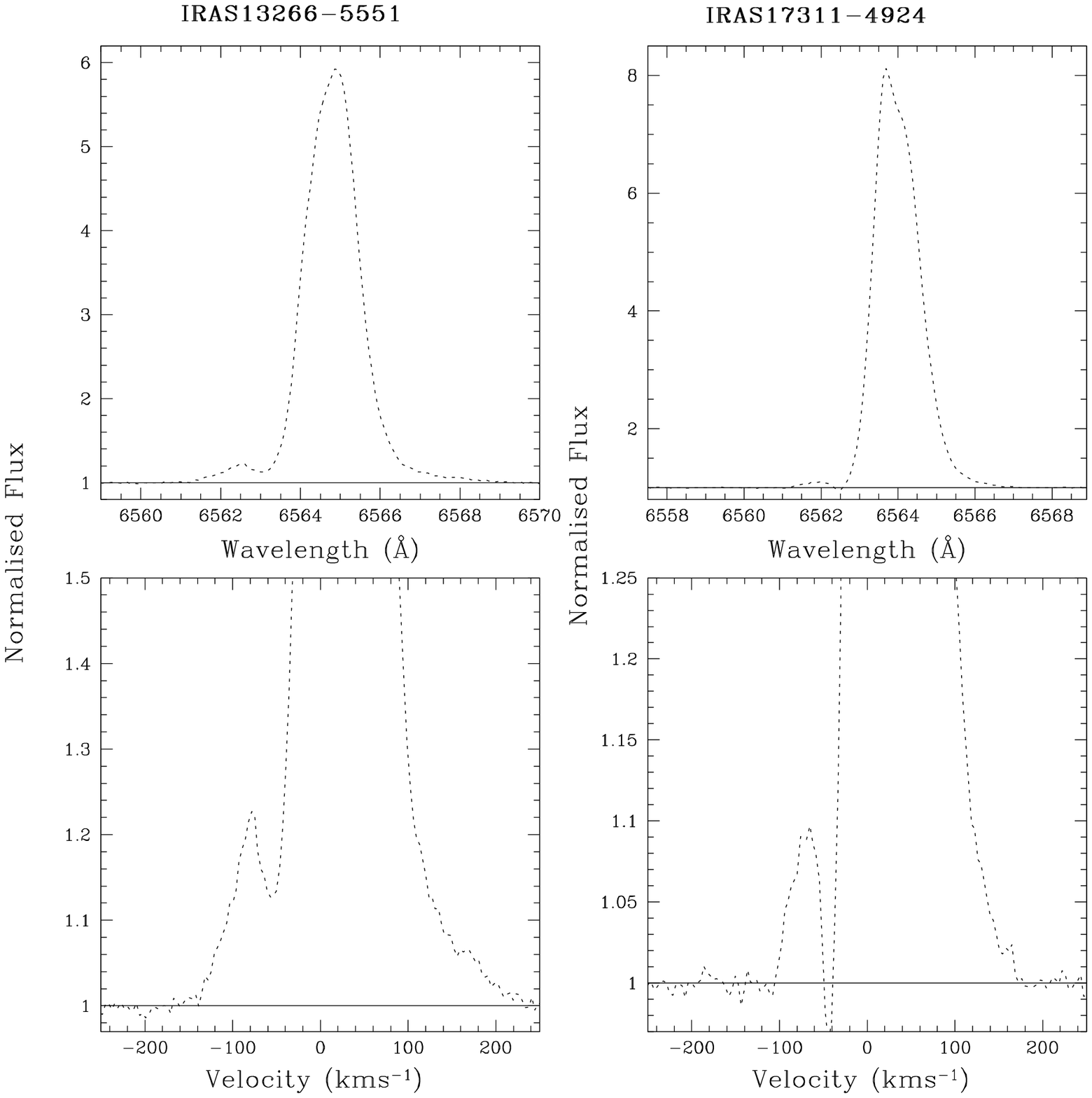}
\caption{The normalised H$_{\alpha}$ profiles (dotted lines) of 
IRAS13266-5551 and IRAS17311-4924}
\end{figure}

\subsection {Expansion velocities}

Nebular expansion velocities were estimated from the
FWHM of the  [O~I] and [N~II] (6548.1 \AA~) emission lines
using V$_{\rm exp}$=0.50 FWHM (Table 4). The mean expansion velocity 
for IRAS17311-4924 from [O~I] lines is 11.99 km s$^{-1}$. 
The higher velocity from [N~II] is due to the fact that 
[N~II] emission dominates in the outermost ionised layers 
of the nebulae and V$_{\rm exp}$ increases with radial distance from the
central stars (see e.g. Weinberger, 1989).

Based on CO observations Loup et al. (1990) and Nyman et al. (1992) 
estimated expansion velocities of 11 km s$^{-1}$ and 14.1 km s$^{-1}$ 
respectively in the case of IRAS17311-4924.   

Radial velocity of absorption component 2 of the Na~I profile 
(18.30 km s$^{-1}$, see Table 3d) in IRAS17311-4924 is comparable 
with the expansion velocity of the star reaffirming its 
possible circumstellar origin.

\subsection{Atmospheric parameters and abundances}

The presence of He~I lines and the absence of He~II lines in 
IRAS13266-5551 and IRAS17311-4924 indicates 
18000~K $\leq$ $T_{\rm eff}$ $\leq$ 25000~K (Miroshnichenko 
et al., 1998). We used Kurucz's WIDTH9 program and 
the spectrum synthesis code, SYNSPEC (Hubeny et al., 1985) along with 
solar metallicity Kurucz (1994) model atmospheres to derive the 
atmospheric parameters and elemental abundances under the LTE approximation. 

The usual criterion for determining the effective temperature
($T_{\rm eff}$), gravity (log $g$) and microturbulent velocity 
($\xi_{\rm t}$) of a star, is to obtain a zero slope respectively
in plots of (i) log abundances for a particular species Vs. 
lower excitation potentials of that species (ii) log 
abundances for two species of a particular element (e.g. Fe~II
and Fe~III) Vs. lower excitation potentials and (iii) log 
abundances for a particular species Vs. equivalent widths. 

The maximum number of absorption lines in the spectrum of IRAS13266-5551
are those of N~II. The majority of the
N~II lines are strong with W$_{\rm \lambda}$ $\geq$ 100 m\AA~. Besides, 
the observed N~II lines fall in a narrow range of lower excitation
potentials (Tables 2a and 3a). This coupled with the lack of 
two ionisation species of any element does not allow us to 
employ the usual criterion for determining $T_{\rm eff}$,
log $g$ and $\xi_{\rm t}$. Hence, for IRAS13266-5551 
we obtained abundances of C~II, N~II, O~II, and Fe~III 
with the WIDTH9 program for various combinations of $T_{\rm eff}$, 
log $g$ and $\xi_{\rm t}$. We covered 18000~K $\leq$ $T_{\rm eff}$ $\leq$ 24000~K 
and 5 km s$^{-1}$ $\leq$ $\xi_{\rm t}$ $\leq$ 45 km s$^{-1}$. From the 
Kurucz (1994) model atmospheres, the log $g$ value was limited
to a minimum of 3.0. For each combination of these parameters, 
we then synthesised the spectrum using SYNSPEC. The best fit to 
the observed spectrum was obtained for 
$T_{\rm eff}$ = 23000~K, log $g$ = 3.0, $\xi_{\rm t}$ = 10 km s$^{-1}$
(Fig. 4).

IRAS17311-4924 has fewer absorption lines. The
maximum number (8) of absorption lines that we find in our
spectrum are those of N~II and Ne~I.
While the N~II lines are strong with W$_{\rm \lambda}$ $\geq$ 100 mA,
the Ne~I lines are very sensitive to NLTE effects (see Sec. 3.8.5
below). Due to the small number of C~II, N~II and O~II lines
and the lack of iron lines, we were unable to estimate the
atmospheric parameters and metallicity. The star suffers significant
circumstellar extinction, E(B$-$V)$_{\rm C.S.}$ = 0.39 
(Gauba and Parthasarathy, 2003) and the circumstellar extinction
law in the UV was found to be linear in ${\rm \lambda^{-1}}$. Hence, the
(B$-$V) color of the star cannot be used for temperature estimation.
An estimate of the temperature and gravity may be made 
from the spectral type of the star. Parthasarathy et al. (2000b) classified
it as B1IIe which corresponds to $T_{\rm eff}$ = 20300~K and log $g$ = 3.0 
(Lang, 1992).

\setcounter{figure}{3}
\begin{figure}
\renewcommand{\thefigure}{\arabic{figure}}
\epsfig{figure=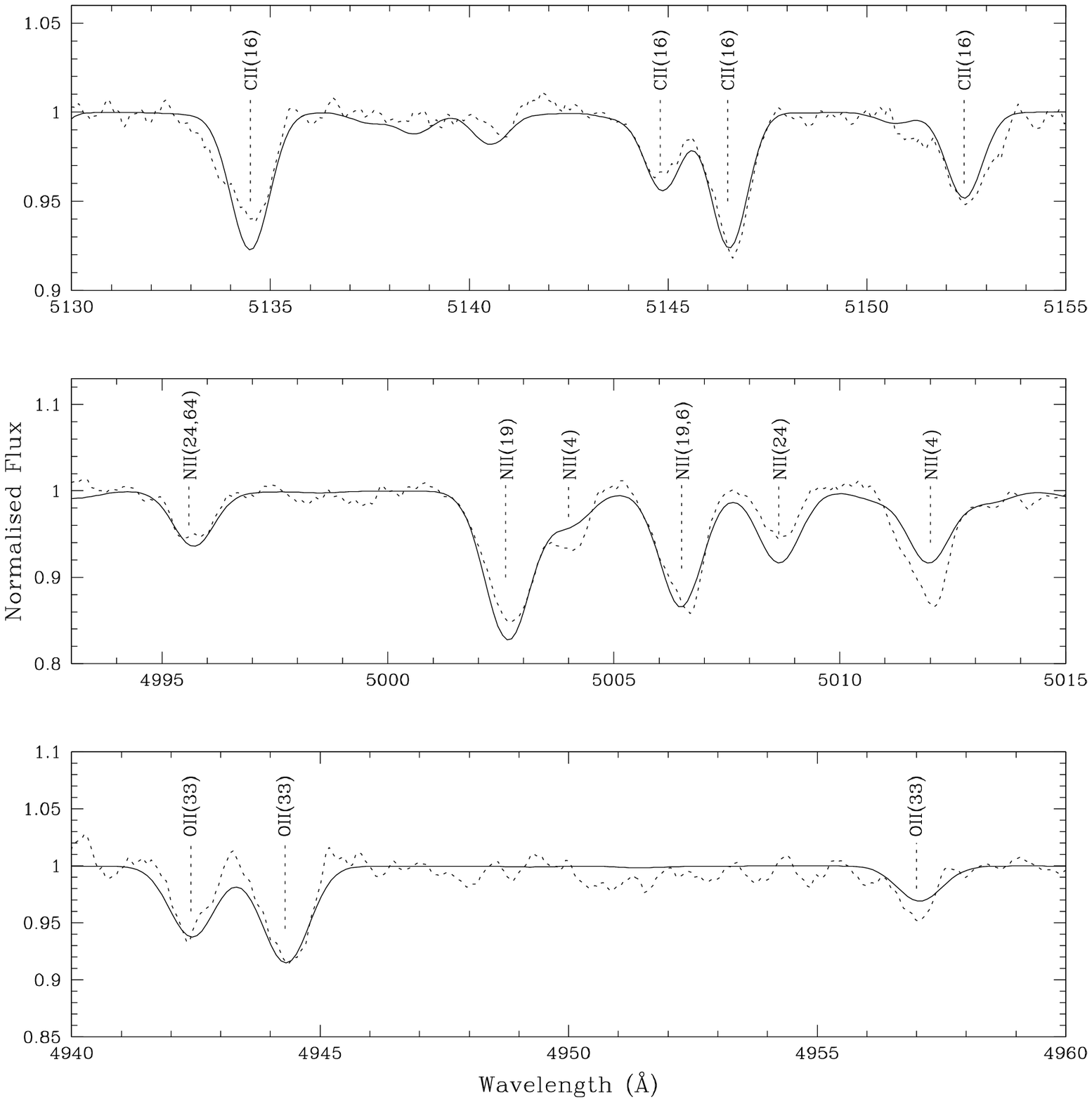}
\caption{Regions showing the C~II, N~II and O~II absorption lines in
the observed spectrum (dotted line) of IRAS13266-5551 
plotted along with the synthetic spectrum (solid line) for 
the derived atmospheric parameters ($T_{\rm eff}$ = 23000~K, 
log $g$ = 3.0, $\xi_{\rm t}$ = 10 km s$^{-1}$) and elemental 
abundances (Table 5a) of the star.}
\end{figure}

\subsubsection{He~I lines} 

We used the 5047.738 \AA~ He~I (47) absorption line in IRAS13266-5551
and estimated the helium abundance in this star (Table 5a). The
estimated abundance is somewhat uncertain since only one line
has been used in the analysis. The helium enrichment may
indicate the evolved nature of the central star.

\subsubsection{C~II lines}

We observed 12 C~II absorption lines in IRAS13266-5551 and
6 C~II absorption lines in IRAS17311-4924. Some of these lines are 
weak and are blended with other lines.
The ISO spectrum of IRAS17311-4924 (Gauba and 
Parthasarathy, 2004) indicated carbon-rich circumstellar dust. 
The presence of carbon lines and their strengths compared to
that of standard stars in the UV (Gauba and Parthasarathy, 2003)
and optical spectra of this star indicate normal or slight 
overabundance of carbon.

\subsubsection{N~II lines}

Since strong lines (W$_{\rm \lambda}$ $\geq$ 100 m\AA~) are usually affected 
by microturbulence, the use of these lines
in determining the atmospheric parameters of the star may 
contribute to systematic errors. Hence N~II lines with 
W$_{\rm \lambda}$ $>$ 100 m\AA~ have been excluded from the abundance 
analysis of IRAS13266-5551.

\subsubsection{O~I triplet and O~II lines}

The equivalent widths of the O~I triplet in the spectra of 
the hot post-AGB stars, LSII$+$34$^{\circ}$26 
(B1.5 Ia, Garc\'ia-Lario et al., 1997; 
Arkhipova et al., 2001b) and IRAS01005+7910 (Klochkova et al., 2002) 
are 0.95 \AA~ and 0.75 \AA~ respectively. The (total) equivalent widths 
of the O~I triplet in IRAS13266-5551 and IRAS17311-4924
are 0.741 \AA~ and 1.479 \AA~ respectively (Tables 2a and 3a). 
The O~I triplet at ${\rm \lambda}$7773 \AA~ is known to be sensitive to 
NLTE effects. For the atmospheric parameters of IRAS13266-5551,
using LTE analysis, we attempted to synthesise the O~I triplet 
with the SYNSPEC code. This required oxygen abundances (log $\epsilon$(O))
in excess of 10.5. In contrast, the derived oxygen abundance using 
LTE analysis for the O~II lines in IRAS13266-5551 is 8.78 (Table 5a). 
Such large discrepancies between the oxygen abundances derived 
from the O~I triplet and the O~II lines has also been observed in the 
hot post-AGB star IRAS01005+7910 (Klochkova et al., 2002). 

The O~II lines at 5161.349 \AA~ (W$_{\rm \lambda}$ = 188.3 m\AA~),
5208.14 (W$_{\rm \lambda}$ = 81.1 m\AA~) and 
6723.222 \AA~ (W$_{\rm \lambda}$ = 168.5 m\AA~) in IRAS13266-5551 
may also have significant NLTE effects. We could not obtain a good
fit to these lines and they were excluded from the abundance analysis.

\subsubsection{Ne~I lines}

The derived neon abundance for IRAS13266-5551 is unusually 
high (Table 5a). However, we could still not obtain a good fit to the 
majority of Ne~I lines in the spectrum of this star. Auer 
and Mihalas (1973) showed that for stars in the range B2 to B5 the neon 
abundance deduced from LTE analyses is systematically in error by about a 
factor of five. They computed equivalent widths of Ne~I lines (${\rm \lambda\lambda}$ 
5852.5 \AA~ to 6598.9 \AA~) for 15000~K $\leq$ $T_{\rm eff}$ $\leq$ 22500~K, 
log $g$ = 3.0 and 4.0 and solar neon abundance. For $T_{\rm eff}$ = 22500~K, 
log $g$ = 3.0, they found that the LTE equivalent widths were almost a factor
of three smaller than the NLTE equivalent widths of these lines.

\subsubsection{Al~III and Si~III lines}

Only two lines of Al~III and one of Si~III were identified in each of the
two stars. These lines are very strong (ref. Tables 2a and 3a) in
both the stars. e.g. in IRAS13266-5551, Si~III, 5741.264 \AA~ has
W$_{\rm \lambda}$ = 392.9 m\AA~. Using WIDTH9 and spectrum synthesis
we derived log $\epsilon$ (Al) = 7.91 $\pm$ 0.34 and 
log $\epsilon$ (Si) = 9.23 for IRAS13266-5551.
However, these abundances may be an overestimate.

\subsubsection{S~II lines}

S~II absorption lines were identified in IRAS13266-5551. 
Since all except one S~II line (5475.025 \AA~) were 
identified as blends, we used the spectrum synthesis code
SYNSPEC to estimate the sulphur abundance in this star
(Table 5a). This value may therefore be treated as an upper
limit.

\subsubsection{Iron lines}

From the Fe~III absorption lines, we estimated [Fe/H] = $-$0.17
in the case of IRAS13266-5551. The iron lines in
IRAS17311-4924 only appear in emission or as P-Cygni profiles. 

\subsubsection{Estimated errors}

The standard deviations ($\sigma$) measure the scatter in the
abundances due to individual lines of a species. Table 5a
gives the value of $\sigma$ for each species as estimated using WIDTH9. The
true error, i.e. the standard deviation of the mean $\sigma$/$\sqrt(n)$,
would be smaller for species with a greater number of lines (n).

Kurucz's solar metallicity models in the range 
18000~K $\leq$ T$_{\rm eff}$ $\leq$ 25000~K
are available only in steps of $\Delta$T$_{\rm eff}$ = $\pm$ 1000~K and
$\Delta$ log $g$ = $\pm$ 0.5. Hence, the temperature (T$_{\rm eff}$)
and gravity (log $g$) of IRAS13266-5551 are estimated to an accuracy of 
$\pm$ 1000~K and $\pm$ 0.5 respectively. Table 5b gives the uncertainites
in the abundances due to uncertainities in the model atmospheric
parameters taking $\Delta$T$_{\rm eff}$ = $+$ 1000~K, $\Delta$log $g$ = $+$ 0.5
and $\Delta\xi_{\rm t}$ = $+$1 km s$^{-1}$. The quadratic sum of the uncertainities 
due to the model parameters is given by $\sigma_{\rm m}$ (Table 5b).

\section{Discussion and conclusions}

From LTE analysis of the high resolution optical spectrum of  
IRAS13266-5551 we find the atmospheric parameters to be
$T_{\rm eff}$ = 23000~K $\pm$ 1000~K, log $g$ =3.0 $\pm$ 0.5 and 
$\xi_{\rm t}$ = 10 km s$^{-1}$.
The lines of Ne~I, a few O~II lines and the O~I triplet
indicate that NLTE effects may be significant in these stars. 
Also, being hot stars it is important to 
obtain observations shortward of 4900 \AA~. The absorption lines 
in the blue may significantly improve our estimates of the stellar 
parameters and elemental abundances. Eg. Mooney et al. (2002), 
Klochkova et al. (2002) and Ryans et al. (2003) have obtained 
observations from $\sim$ 3700 \AA~ onwards. Recently, from NLTE analysis of 
the absorption lines in the blue region, Ryans et al. (2003) estimated
the atmospheric parameters of two hot post-AGB stars,
IRAS18062+2410 (SAO85766) and IRAS19590-1249 (LSIV-12$^{\circ}$111).
A similar analysis is required for IRAS13266-5551 and IRAS17311-4924. 
Hence, we would like to emphasise that our LTE analysis based on
lines longward of 4900 \AA~ is only a first approximation.

We estimated heliocentric radial velocities (V$_{\rm r}$) of 
$+$65.31 $\pm$ 0.34 km s$^{-1}$  and $+$27.55 $\pm$ 0.74 km s$^{-1}$
for IRAS13266-5551 and IRAS17311-4924 respectively. 

Preliminary estimates of the CNO abundances in IRAS13266-5551  
indicate that these elements are overabundant with
[C/Fe]=$+$0.45, [N/Fe]=$+$0.48 and [O/Fe]=$+$0.26 suggesting that the
products of helium burning have been brought to the surface as a result of 
third dredge-up on the AGB. A comparison with average CNO abundances
for mainsequence B-stars from the Ori OB1 association (Table 5a; Kilian, 1992)
also indicates that IRAS13266-5551 is an evolved star 
and has gone through the dredge-up episodes during its evolution.
We could not estimate the atmospheric parameters and chemical abundances
for IRAS17311-4924. 

McCausland et al. (1992) and Conlon et al. (1993b) derived the 
chemical composition of several high
galactic latitude hot post-AGB stars. In addition to being metal-poor, these 
stars showed severe carbon deficiency. Similar carbon depletions were also
reported in other hot post-AGB stars at high galactic latitudes 
e.g. LSII$+$34$^{\circ}$26 (Garc\'ia-Lario et al., 1997),
PG1323-086 and PG1704+222 (Moehler and Heber, 1998) and SAO85766
(Parthasarathy et al., 2000b). In contrast, the hot post-AGB star,
IRAS01005+7910 (Klochkova et al., 2002) was found to be
carbon-rich. For, IRAS13266-5551 we estimated C/O $\sim$ 0.78.

Finally, from our optical spectra we conclude that 
IRAS13266-5551 and IRAS17311-4924 are most
likely in the post-AGB phase of evolution. These stars are unlikely
to be luminous blue variables (LBVs). Their spectra are
very similar to the hot post-AGB stars IRAS18062+2410 
(SAO85766; Parthasarathy et al., 2000b) and IRAS01005+7910 
(Klochkova et al., 2002). LBVs are usually found in the galactic 
plane and are often associated with star forming regions. IRAS13266-5551
and IRAS17311-4924 on the other hand, are
at high galactic latitudes and are not associated with any star forming
region. Further, whereas, LBVs are characterised by large amplitude light
variations, these stars may show small amplitude, irregular 
light variations similar to that found in the high galactic latitude
rapidly evolving hot post-AGB star, SAO85766 (Arkhipova, 1999, 2000).
Photometric monitoring of these stars is required. 

\setcounter{table}{1}
\begin{table}
\begin{center}
\renewcommand{\thetable}{\arabic{table}a}
\caption{Absorption lines in IRAS13266-5551 (CPD-55 5588)}
\begin{tabular}{|c|c|c|c|c|c|c|c|}
\hline
${\rm \lambda}_{\rm obs.}$ & ${\rm \lambda}_{\rm lab.}$ & Ident. & W$_{\rm \lambda}$ &
log (gf) & $\chi$ & $\Delta {\rm \lambda}$ & V$_{\rm r}$ \\
   (\AA~)            &   (\AA~)             &        &   (\AA~)       &
         &  (eV)  &      (\AA~)      & km s$^{-1}$   \\
\hline \hline
4942.394  & 4941.105      & O~II (33)     & 0.0498     &    0.080 & 26.55$-$29.06 & 1.289 & $+$63.03 \\
4944.33   & 4943.003      & O~II (33)     & 0.0901     &    0.370 & 26.56$-$29.07 & 1.327 & $+$65.31 \\
4957.005  & 4955.738      & O~II (33)     & 0.0437     & $-$0.420 & 26.56$-$29.06 & 1.267 & $+$61.47 \\
4995.447  & 4994.360      & N~II (24, 64) & 0.028      & $-$0.080 & 20.94$-$23.42 &       &          \\ 
5002.728  & 5001.474      & N~II (19)     & 0.222      &    0.450 & 20.65$-$23.13 &       & blend    \\
5003.995  & 5002.703      & N~II (4)      & 0.0665     & $-$1.020 & 18.46$-$20.94 &       & blend    \\
5006.535  & 5005.150      & N~II (19, 6)  & 0.1499     &    0.610 & 20.66$-$23.14 &       &          \\ 
5008.656  & 5007.328      & N~II (24)     & 0.0494     &    0.160 & 20.94$-$23.41 & 1.328 & $+$64.33 \\ 
5011.923  & 5010.621      & N~II (4)      & 0.1593     & $-$0.520 & 18.46$-$20.94 & 1.302 & $+$62.72 \\
5046.411  & 5045.099      & N~II (4)      & 0.1784     & $-$0.330 & 18.48$-$20.94 & 1.312 & $+$62.79 \\
5049.117  & 5047.738      & He~I (47)     & 0.1329     & $-$1.600 & 21.22$-$23.67 & 1.379 & $+$66.73 \\
5075.435  & 5073.592      & N~II (10)     &            & $-$1.280 & 18.49$-$20.94 &       &          \\
          & $+$ 5073.903  & Fe~III (5)    &            & $-$2.557 &  8.65$-$11.09 &       &          \\
5088.355  & 5086.701      & Fe~III (5)    & 0.0372     & $-$2.590 &  8.65$-$11.09 & 1.654 & $+$82.32 \\ 
5129.21   & 5127.387      & Fe~III (5)    & 0.0390     & $-$2.218 &  8.65$-$11.07 & 1.823 & $+$91.43 \\
5134.133  & 5132.947      & C~II (16)     & 0.102      & $-$0.240 & 20.70$-$23.12 &       &          \\
          & $+$ 5133.281  & C~II (16)     &            & $-$0.200 & 20.70$-$23.12 &       &          \\ 
5144.795  & 5143.494      & C~II (16)     & 0.045      & $-$0.240 & 20.70$-$23.11 & 1.301 & $+$60.65 \\ 
5146.582  & 5145.165      & C~II (16)     & 0.0896     &    0.160 & 20.71$-$23.12 & 1.417 & $+$67.39 \\
5152.49   & 5151.085      & C~II (16)     & 0.0781     & $-$0.200 & 20.71$-$23.12 & 1.405 & $+$66.60 \\
5157.759  & 5156.111      & Fe~III (5)    & 0.0398     & $-$2.018 &  8.64$-$11.04 & 1.648 & $+$80.66 \\
5161.349  & 5160.026      & O~II (32)     & 0.1883     & $-$0.660 & 26.56$-$28.96 & 1.323 & $+$61.69 \\
5208.14   & 5206.715      & O~II (32)     & 0.0811     & $-$0.860 & 26.56$-$28.94 & 1.425 & $+$66.87 \\ 
5220.938  &               & UN            & 0.133      &          &               &       &          \\
5455.271  & 5453.790      & S~II (6)      & 0.1352     &    0.560 & 13.67$-$15.94 &       &          \\ 
          & $+$ 5454.215  & N~II (29)     &            & $-$0.740 & 21.15$-$23.42 &       &          \\ 
5475.025  & 5473.602      & S~II (6)      & 0.0982     & $-$0.120 & 13.58$-$15.85 & 1.423 & $+$62.76 \\ 
5497.073  & 5495.655      & N~II (29)     & 0.0767     & $-$0.170 & 21.16$-$23.42 & 1.418 & $+$62.18 \\
5641.593  & 5639.980      & S~II (14)     & 0.0480     &    0.330 & 14.07$-$16.26 &       &          \\ 
          & $+$ 5640.549  & C~II (15)     &            & $-$0.750 & 20.70$-$22.90 &       &          \\
5648.699  & 5646.979      & S~II (14)     & 0.0426     &    0.110 & 14.00$-$16.20 &       &          \\
          & $+$ 5648.070  & C~II (15)     &            & $-$0.450 & 20.70$-$22.90 &       &          \\
5668.134  & 5666.629      & N~II (3)      & 0.2938     &    0.010 & 18.47$-$20.65 & 1.505 & $+$64.45 \\
5677.53   & 5676.017      & N~II (3)      & 0.2524     & $-$0.340 & 18.46$-$20.65 & 1.513 & $+$64.74 \\
5681.04   & 5679.558      & N~II (3)      & 0.3913     &    0.280 & 18.48$-$20.67 & 1.482 & $+$63.05 \\
5687.75   & 5686.213      & N~II (3)      & 0.1746     & $-$0.470 & 18.47$-$20.65 & 1.537 & $+$65.86 \\
5698.092  & 5696.603      & Al~III (2)    & 0.3473     &    0.230 & 15.64$-$17.82 & 1.489 & $+$63.19 \\
5712.274  & 5710.766      & N~II (3)      & 0.1726     & $-$0.470 & 18.48$-$20.65 & 1.508 & $+$63.99 \\
5724.268  & 5722.730      & Al~III (2)    & 0.2174     & $-$0.070 & 15.64$-$17.81 & 1.538 & $+$65.39 \\
5741.264  & 5739.734      & Si~III (4)    & 0.3929     & $-$0.160 & 19.72$-$21.88 & 1.53  & $+$64.74 \\
\hline
\end{tabular}
\end{center}
\end{table}

\setcounter{table}{1}
\begin{table}
\begin{center}
\renewcommand{\thetable}{\arabic{table}a}
\caption{contd...}
\begin{tabular}{|c|c|c|c|c|c|c|c|}
\hline
${\rm \lambda}_{\rm obs.}$ & ${\rm \lambda}_{\rm lab.}$ & Ident. & W$_{\rm \lambda}$ &
log (gf) & $\chi$ & $\Delta {\rm \lambda}$ & V$_{\rm r}$ \\
   (\AA~)            &   (\AA~)             &        &   (\AA~)       &
         &  (eV)  &      (\AA~)      & km s$^{-1}$   \\
\hline \hline
5780.631 & 5780.410     & DIB          & 0.1696     &          &               & 0.221 & $-$3.76  \\
5797.34  & 5797.030     & DIB          & 0.0998     &          &               & 0.310 & $+$0.81  \\
5835.532 & 5833.938     & Fe~III (114) & 0.1198     &    0.616 & 18.51$-$20.63 & 1.594 & $+$66.74 \\
5933.462 & 5931.782     & N~II (28)    & 0.0428     &    0.050 & 21.15$-$23.24 & 1.68  & $+$69.74 \\
5943.301 & 5941.654     & N~II (28)    & 0.0756     &    0.320 & 21.16$-$23.25 & 1.647 & $+$67.93 \\
6144.637 & 6143.063     & Ne~I (1)     & 0.0606     & $-$0.350 & 16.62$-$18.64 & 1.574 & $+$61.64 \\
6196.062 & 6195.990     & DIB          & 0.0214     &          &               & 0.072 & $-$11.74 \\
6203.327 & 6203.060     & DIB          & 0.0221     &          &               & 0.267 & $-$2.32  \\
6381.399 & 6379.617     & N~II (2)     & 0.0587     & $-$0.920 & 18.47$-$20.41 & 1.782 & $+$68.57 \\
6403.899 & 6402.246     & Ne~I (1)     & 0.1314     &    0.360 & 16.62$-$18.55 & 1.653 & $+$62.23 \\ 
6483.787 & 6482.049     & N~II (8)     & 0.2212     & $-$0.160 & 18.50$-$20.41 & 1.738 & $+$65.21 \\
6508.432 & 6506.528     & Ne~I (3)     & 0.0275     &    0.030 & 16.67$-$18.58 & 1.904 & $+$72.56 \\
6579.07  & 6578.052$^{\dagger}$ & C~II (2) & 0.0855 &    0.120 & 14.45$-$16.33 & 1.018 & $+$46.43 \\ 
6613.856 & 6613.630     & DIB          & 0.0832     &          &               & 0.226 & $-$4.98 \\
6642.826 & 6640.994     & O~II (4)     & 0.0825     & $-$0.890 & 23.42$-$25.29 & 1.832 & $+$67.53 \\
6723.222 & 6721.358     & O~II (4)     & 0.1685     & $-$0.590 & 23.44$-$25.29 & 1.864 & $+$67.97 \\
6781.99  & 6779.942     & C~II (14)    & 0.0355     &    0.040 & 20.70$-$22.53 &       &          \\
         & $+$ 6780.595 & C~II (14)    &            & $-$0.360 & 20.70$-$22.53 &       &          \\
6785.831 & 6783.907     & C~II (14)    & 0.0458     &    0.320 & 20.71$-$22.54 & 1.924 & $+$69.85 \\
         & 6787.210     & C~II (14)    &            & $-$0.360 & 20.70$-$22.53 &       & weak     \\
         & 6791.465     & C~II (14)    &            & $-$0.250 & 20.70$-$22.53 &       & weak     \\
	 & 6800.687     & C~II (14)    &            & $-$0.330 & 20.70$-$22.54 &       & weak     \\
7034.428 & 7032.413     & Ne~I (1)     & 0.0485     & $-$0.250 & 16.62$-$18.38 & 2.015 & $+$70.73 \\
7773.521 & 7771.944     & O~I (1)      & 0.273      &    0.320 &  9.14$-$10.74 &       & blend    \\
7776.411 & 7774.166     & O~I (1)      & 0.396      &    0.170 &  9.14$-$10.74 &       & blend    \\
7777.68  & 7775.388     & O~I (1)      & 0.072      & $-$0.050 &  9.14$-$10.74 &       & weak     \\

\hline
\end{tabular}

\noindent \parbox{14cm}{Radial velocity of the C~II(2) 6578.052 \AA~ absorption line 
indicates that it may have a P-Cygni profile with a weak emission component, similar
to C~II(2) 6582.882 \AA~ (see Table 2c).}

\end{center}
\end{table}

\setcounter{table}{1}
\begin{table}
\begin{center}
\renewcommand{\thetable}{\arabic{table}b}
\caption{Emission lines in IRAS13266-5551 (CPD-55 5588)}
\begin{tabular}{|c|c|c|c|c|c|c|c|}
\hline
${\rm \lambda}_{\rm obs.}$ & ${\rm \lambda}_{\rm lab.}$ & Ident. & W$_{\rm \lambda}$ &
log (gf) & $\chi$ & $\Delta {\rm \lambda}$ & V$_{\rm r}$ \\
   (\AA~)            &   (\AA~)             &        &   (\AA~)       &
         &  (eV)  &      (\AA~)      & km s$^{-1}$  \\
\hline \hline
          &   5015.678$^{\dagger}$ & He~I (4)     &        & $-$0.820  & 20.62$-$23.09 &       &          \\
5042.128  &   5041.024             & Si~II (5)    & 0.0913 &    0.290  & 10.07$-$12.52 & 1.104 & $+$50.47 \\
5057.172  &   5055.984             & Si~II (5)    & 0.1503 &    0.590  & 10.07$-$12.52 & 1.188 & $+$55.26 \\
5199.059  &                        & UN           & 0.0281 &           &               &       &          \\
5201.533  &                        & UN           & 0.0263 &           &               &       &          \\
5467.921  &   5466.55              & S~II (11)    & 0.0346 &           & 13.62$-$15.88 & 1.371 & $+$60.01 \\ 
5516.046  &                        & UN           & 0.0924 &           &               &       &          \\  
5921.673  &   5920.124             & Fe~III (115) & 0.0420 & $-$0.034  & 18.78$-$20.88 & 1.549 & $+$63.26 \\
5955.113  &   5953.613             & Fe~III (115) & 0.0371 &    0.186  & 18.79$-$20.87 & 1.5   & $+$60.35 \\
5958.987  &   5957.559             & Si~II (4)    & 0.1042 & $-$0.300  & 10.07$-$12.15 & 1.428 & $+$56.68 \\
5980.413  &   5978.90              & Fe~III (117) & 0.3632 &           & 18.73$-$20.80 &       &          \\
          &$+$5978.930             & Si~II (4)    &        &    0.000  & 10.07$-$12.15 &       &          \\
6000.962  &   5999.30              & Fe~III (117) & 0.0328 &           & 18.73$-$20.79 & 1.662 & $+$67.88 \\
6033.913  &   6032.30              & Fe~III (117) & 0.0666 &           & 18.73$-$20.78 & 1.613 & $+$64.99 \\
6096.85   &   6095.37              & C~II (24)    & 0.0226 &           & 22.47$-$24.50 & 1.48  & $+$57.61 \\
6100.046  &   6098.62              & C~II (24)    & 0.0280 &           & 22.47$-$24.50 & 1.426 & $+$54.92 \\
6227.758  &   6226.130             & Al~II (10)   & 0.0100 &    0.050  & 13.07$-$15.06 & 1.628 & $+$63.21 \\
6233.21   &   6231.718             & Al~II (10)   & 0.0421 &    0.400  & 13.07$-$15.06 & 1.492 & $+$56.6  \\
6241.053  &                        & UN           & 0.0398 &           &               &       &          \\
6244.939  &   6243.355             & Al~II (10)   & 0.0281 &    0.670  & 13.08$-$15.06 & 1.584 & $+$60.88 \\
6300.392  &                        & [O~I ] (1F) (atmos.) & 0.0728 &   &               &       &          \\
6348.553  &   6347.091             & Si~II (2)    & 0.2767 &    0.300  &  8.12$-$10.07 & 1.462 & $+$53.87 \\
6363.881  &                        & [O~I] (1F) (atmos.)  & 0.0222 &   &               &       &          \\
6372.811  &   6371.359             & Si~II (2)    & 0.1369 &    0.000  &  8.12$-$10.07 & 1.452 & $+$53.14 \\
6463.481  &                        & atmos.       & 0.0399 &           &               &       &          \\
6547.435  &   6545.80              & Mg~II (23)   & 0.0303 &           & 11.58$-$13.47 & 1.635 & $+$59.70 \\
6549.576  &   6548.1               & [N~II] (1F)  & 0.0209 &           &               & 1.476 & $+$52.39 \\
          &   6583.6$^{*}$         & [N~II] (1F)  &        &           &               &       &          \\
6732.171  &                        & UN           & 0.0259 &           &               &       &          \\
6853.447  &                        & UN           & 0.0463 &           &               &       &          \\
7043.729  &   7042.048             & Al~II (3)    & 0.0818 &    0.350  & 11.32$-$13.08 & 1.681 & $+$56.38 \\
7058.222  &   7056.612             & Al~II (3)    & 0.0619 &    0.130  & 11.32$-$13.07 & 1.61  & $+$53.22 \\
7063.996  &                        & UN           & 0.016  &           &               &       &          \\
7065.332  &   7063.642             & Al~II (3)    & 0.0471 & $-$0.350  & 11.32$-$13.07 & 1.68  & $+$56.12 \\
7067.219  &   7065.188             & He~I (10)    & 0.5544 &           & 20.87$-$22.62 &       &          \\
          &$+$7065.719             & He~I (10)    &        &           & 20.87$-$22.62 &       &          \\ 
7233.203  &   7231.332             & C~II (3)     & 0.316  &    0.070  & 16.33$-$18.05 & 1.871 & $+$62.39 \\
7238.37   &   7236.421             & C~II (3)     & 0.488  &    0.330  & 16.33$-$18.05 & 1.949 & $+$65.57 \\
          &   7281.349$^{\dagger}$ & He~I (45)    &        & $-$0.840  & 21.22$-$22.92 &       &          \\
\hline
\end{tabular}

\noindent \parbox{14cm}{$^\dagger$ : the He~I(4) 5015.675 \AA~ and He~I(45) 7281.349 \AA~
emission lines are superposed on the corresponding absorption profiles of these lines.
The asymmetric nature of these emission lines suggests that they may have P-Cygni
profiles. $^{*}$ : [N~II] (1F) 6583.6 \AA~ is blended with the emission component 
of C~II (2) 6582.882 \AA~ P-Cygni profile.}

\end{center}
\end{table}

\setcounter{table}{1}
\begin{table}
\begin{center}
\renewcommand{\thetable}{\arabic{table}b}
\caption{contd...}
\begin{tabular}{|c|c|c|c|c|c|c|c|}
\hline
${\rm \lambda}_{\rm obs.}$ & ${\rm \lambda}_{\rm lab.}$ & Ident. & W$_{\rm \lambda}$ &
log (gf) & $\chi$ & $\Delta {\rm \lambda}$ & V$_{\rm r}$ \\
   (\AA~)            &   (\AA~)             &        &   (\AA~)       &
         &  (eV)  &      (\AA~)      & km s$^{-1}$   \\
\hline \hline

7316.384  &                        & atmos.       & 0.0237 &           &               &       & \\         
7379.562  &                        & UN           & 0.0736 &           &               &       & \\
7413.63   &                        & atmos.       & 0.0258 &           &               &       & \\
7464.238  &                        & UN           & 0.0624 &           &               &       & \\ 
7468.316  &                        & atmos.       & 0.0306 &           &               &       & \\
7497.305  &                        & UN           & 0.0161 &           &               &       & \\
7504.157  &                        & UN           & 0.0123 &           &               &       & \\ 
7514.875  &                        & UN           & 0.0569 &           &               &       & \\ 
7564.141  &                        & UN           & 0.0178 &           &               &       & \\ 
7712.676  &                        & atmos.       & 0.0276 &           &               &       & \\
7717.028  &                        & atmos.       & 0.0126 &           &               &       & \\
7750.781  &                        & atmos.       & 0.0322 &           &               &       & \\
7794.201  &                        & atmos.       & 0.0223 &           &               &       & \\
7821.623  &                        & atmos.       & 0.0517 &           &               &       & \\ 
7851.066  &                        & UN           & 0.1470 &           &               &       & \\     
7853.381  &                        & atmos.       & 0.0263 &           &               &       & \\
7854.82   &                        & UN           & 0.0248 &           &               &       & \\
7878.902  &    7877.054            & Mg~II (8)    & 0.2393 &    0.390  & 9.99$-$11.57  & 1.848 & $+$55.15 \\
7898.179  &    7896.367            & Mg~II (8)    & 0.3963 &    0.650  & 10.00$-$11.57 & 1.812 & $+$53.61 \\
7913.766  &                        & atmos.       & 0.0636 &           &               &       & \\  
7921.164  &                        & atmos.       & 0.0420 &           &               &       & \\ 
7964.76   &                        & atmos.       & 0.0554 &           &               &       & \\
7993.434  &                        & atmos.       & 0.0468 &           &               &       & \\   
8001.991  &    8000.12             & [Cr~II] (1F) & 0.0441 &           &               & 1.871 & $+$54.93 \\  
8025.808  &                        & atmos.       & 0.0350 &           &               &       & \\
8062.346  &                        & atmos.       & 0.0199 &           &               &       & \\ 
8215.857  &                        & UN           & 0.0891 &           &               &       & \\ 
8236.683  &                        & UN           & 0.2139 &           &               &       & \\
\hline
\end{tabular}

\end{center}
\end{table}

\setcounter{table}{1}
\begin{table}
\begin{center}
\renewcommand{\thetable}{\arabic{table}c}
\caption{Lines with P-Cygni profiles in IRAS13266-5551 (CPD-55 5588). Equivalent
widths of the absorption and emission components of the P-Cygni profiles are 
given. Wind velocities are estimated from the blue absorption edges of the
P-Cygni profiles.}
\begin{tabular}{|c|c|c|c|c|c|c|}
\hline
${\rm \lambda}_{\rm lab}$ & Ident. & W$_{\rm \lambda}$ (absorption) & W$_{\rm \lambda}$ (emission) & log (gf) &
$\chi$ & Wind Velocity\\
       (\AA~)       &        &  (\AA~)                    & (\AA~)                   &          &
(eV) & km s$^{-1}$\\ 
\hline \hline
5875.618     & He~I (11)     & 0.079 & 0.484 &    0.410 & 20.97$-$23.08 & \\
$+$5875.650  & He~I (11)     &       &       & $-$0.140 & 20.97$-$23.08 & \\
$+$5875.989  & He~I (11)     &       &       & $-$0.210 & 20.97$-$23.08 & \\
6562.817     & H$_{\alpha}$  &       & 6.902 &    0.710 & 10.15$-$12.04 & \\
6582.882$^{\dagger}$ & C~II (2) & 0.031 &    & $-$0.180 & 14.45$-$16.33 & \\
6678.149     & He~I (46)     & 0.212 & 0.076 &    0.330 & 21.22$-$23.08 & $-$101.34\\

\hline
\end{tabular}

\noindent \parbox{14cm}{Owing to the broad wings of the H$_{\alpha}$ emission
component, the absorption component of the P-Cygni profile lies above
the normalised continuum (see Fig. 3). $^{\dagger}$ : the emission component 
of C~II(2) 6582.882 \AA~ P-Cygni profile is blended with [N~II](1F) 6583.6 \AA~}

\end{center}
\end{table}

\setcounter{table}{1}
\begin{table}
\begin{center}
\renewcommand{\thetable}{\arabic{table}d}
\caption{Absorption (a) and emission (e) components of Na~I D$_{2}$ (5889.953 \AA~)
and Na~I D$_{1}$ (5895.923 \AA~) lines in the spectrum of IRAS13266-5551 (CPD-55 5588).
W$_{\rm \lambda}$ are the equivalent widths of the components and V$_{\rm r}$ are the
respecitve heliocentric radial velocities.}
\begin{tabular}{|c|c|c|c|c|c|c|c|}
\hline
           & & \multicolumn{6}{c|}{IRAS13266-5551} \\ \cline{3-8}
          & & \multicolumn{3}{c|}{Na~I D$_{2}$} & \multicolumn{3}{c|}{Na~I D$_{1}$} \\ \cline{3-8}
Component & & ${\rm \lambda}_{\rm obs.}$ & W$_{\rm \lambda}$ & V$_{\rm r}$ &
${\rm \lambda}_{\rm lab}$ & W$_{\rm \lambda}$ & V$_{\rm r}$ \\
          & & (\AA~) & (\AA~) & (km s$^{-1}$) & (\AA~) & (\AA~) & (km s$^{-1}$) \\
\hline
1. & a & 5889.54  & 0.168 & $-$36.28 & 5895.432 & 0.062 & $-$40.22 \\
2. & a & 5890.003 & 0.551 & $-$12.69 & 5896.002 & 0.415 & $-$11.22 \\
3. & a & 5890.366 & 0.292 & $+$5.80  & 5896.38  & 0.203 & $+$8.01 \\
4. & a & 5890.938 & 0.021 & $+$34.92 & 5896.913 & 0.029 & $+$35.13 \\
5. & e & 5891.324 & 1.371 & $+$54.59 & 5897.373 & 0.007 & $+$58.54 \\

\hline 
\end{tabular}
\end{center}
\end{table}

\setcounter{table}{2}
\begin{table}
\begin{center}
\renewcommand{\thetable}{\arabic{table}a}
\caption{Absorption lines in IRAS17311-4924 (Hen3-1428)}
\begin{tabular}{|c|c|c|c|c|c|c|c|}
\hline
${\rm \lambda}_{\rm obs.}$ & ${\rm \lambda}_{\rm lab.}$ & Ident. & W$_{\rm \lambda}$ &
log (gf) & $\chi$ & $\Delta {\rm \lambda}$ & V$_{\rm r}$ \\
   (\AA~)            &   (\AA~)             &        &   (\AA~)       &
         &  (eV)  &      (\AA~)      & km s$^{-1}$   \\
\hline \hline
5045.576             & 5044.8       & C~II (35)   & 0.0976  &          & 24.27$-$26.71 &            &         \\ 
                     & $+$5045.099  & N~II (4)    &         & $-$0.330 & 18.48$-$20.94 &            &         \\ 
5047.665             & 5047.2       & C~II (35)   & 0.0686  & $-$1.00  & 24.27$-$26.71 &    0.465   & $+$25.17 \\ 
5133.486             & 5132.947     & C~II (16)   & 0.0643  & $-$0.240 & 20.70$-$23.12 &            &          \\
                     & $+$5133.281  & C~II (16)   &         & $-$0.200 & 20.70$-$23.12 &            &          \\
5144.009             & 5143.494     & C~II (16)   & 0.0436  & $-$0.240 & 20.70$-$23.12 &    0.515   & $+$27.57 \\
5145.53              & 5145.011     & Ne~I (34)   & 0.0913  &          & 18.62$-$21.02 &            &          \\
                     & $+$5145.165  & C~II (16)   &         &    0.160 & 20.71$-$23.12 &            &          \\
5151.433             & 5151.085     & C~II (16)   & 0.0294  & $-$0.200 & 20.71$-$23.12 &    0.348   & $+$17.8  \\
5160.659$^{\dagger}$ & 5160.026     & O~II (32)   & 0.0584  & $-$0.660 & 26.55$-$28.96 &            & blend    \\
5207.288             & 5206.715     & O~II (32)   & 0.0137  & $-$0.860 & 26.56$-$28.94 &    0.573   & $+$30.54 \\    
5496.109             & 5495.655     & N~II (29)   & 0.0215  & $-$0.170 & 21.16$-$23.42 &    0.454   & $+$22.31 \\ 
5667.181             & 5666.629     & N~II (3)    & 0.1652  &    0.010 & 18.46$-$20.65 &    0.552   & $+$26.75 \\
5676.614             & 5676.017     & N~II (3)    & 0.1386  & $-$0.340 & 18.46$-$20.65 &    0.597   & $+$29.08 \\
5680.022             & 5679.558     & N~II (3)    & 0.2475  &    0.280 & 18.48$-$20.67 &    0.464   & $+$22.04 \\
5686.792             & 5686.213     & N~II (3)    & 0.1017  & $-$0.470 & 18.47$-$20.65 &    0.579   & $+$28.08 \\
5697.17              & 5696.603     & Al~III (2)  & 0.1941  &    0.230 & 15.64$-$17.82 &    0.567   & $+$27.39 \\
5711.247             & 5710.766     & N~II (3)    & 0.0758  & $-$0.470 & 18.48$-$20.65 &    0.481   & $+$22.80 \\
5723.368             & 5722.730     & Al~III (2)  & 0.1158  & $-$0.070 & 15.64$-$17.81 &    0.638   & $+$30.97 \\
5740.416             & 5739.734     & Si~III (4)  & 0.1923  & $-$0.160 & 19.72$-$21.88 &    0.682   & $+$33.18 \\
5780.096             & 5780.410     & DIB         & 0.1099  &          &               & $-$0.314   & $-$18.77 \\
6143.323             & 6143.063     & Ne~I (1)    & 0.1289  & $-$0.350 & 16.62$-$18.64 &    0.26    & $+$10.23 \\
6163.944             & 6163.594     & Ne~I (5)    & 0.0460  & $-$0.590 & 16.72$-$18.73 &    0.35    & $+$14.56 \\
6266.823             & 6266.495     & Ne~I (5)    & 0.0351  & $-$0.530 & 16.72$-$18.69 &    0.328   & $+$13.23 \\
6334.755             & 6334.428     & Ne~I (1)    & 0.0709  & $-$0.310 & 16.62$-$18.58 &    0.327   & $+$13.02 \\
6380.267             & 6379.617     & N~II (2)    & 0.0216  & $-$0.920 & 18.47$-$20.41 &    0.65    & $+$28.10 \\
6383.238             & 6382.991     & Ne~I (3)    & 0.0587  & $-$0.260 & 16.67$-$18.61 &    0.247   & $+$9.14  \\
6402.463             & 6402.246     & Ne~I (1)    & 0.2416  &    0.360 & 16.62$-$18.56 &    0.217   & $+$7.7   \\
6482.705             & 6482.049     & N~II (8)    & 0.1270  & $-$0.160 & 18.50$-$20.41 &    0.656   & $+$27.89 \\
6506.754             & 6506.528     & Ne~I (3)    & 0.0441  &    0.030 & 16.67$-$18.58 &    0.226   & $+$7.95  \\
6641.812             & 6640.994     & O~II (4)    & 0.0738  & $-$0.890 & 23.42$-$25.29 &    0.818   & $+$34.48 \\
6722.179             & 6721.358     & O~II (4)    & 0.1031  & $-$0.590 & 23.44$-$25.29 &    0.821   & $+$34.17 \\
7032.735             & 7032.413     & Ne~I (1)    & 0.1089  & $-$0.250 & 16.62$-$18.38 &    0.322   & $+$11.27 \\
7772.134             & 7771.944     & O~I (1)     & 0.743   &    0.320 &  9.14$-$10.74 &            & blend \\
7774.474             & 7774.166     & O~I (1)     & 0.530   &    0.170 &  9.14$-$10.74 &            & blend \\
7775.758             & 7775.388     & O~I (1)     & 0.206   & $-$0.050 &  9.14$-$10.74 &            & blend \\

\hline
\end{tabular}

\noindent \parbox{14cm}{$^{\dagger}$ : O~II (32) 5160.026 \AA~ absorption line is blended with
[Fe~II] (19F) 5158.81 \AA~.}

\end{center}
\end{table}

\setcounter{table}{2}
\begin{table}
\begin{center}
\renewcommand{\thetable}{\arabic{table}b}
\caption{Emission lines in IRAS17311-4924 (Hen3-1428)}
\begin{tabular}{|c|c|c|c|c|c|c|c|}
\hline
${\rm \lambda}_{\rm obs.}$ & ${\rm \lambda}_{\rm lab.}$ & Ident. & W$_{\rm \lambda}$ &
log (gf) & $\chi$ & $\Delta {\rm \lambda}$ & V$_{\rm r}$ \\
   (\AA~)            &   (\AA~)             &        &   (\AA~)       &
         &  (eV)  &      (\AA~)      & km s$^{-1}$ \\
\hline \hline

5041.596 & 5041.024       & Si~II (5)     & 0.0952 &    0.290 & 10.07$-$12.53 & 0.572 & $+$31.57 \\
5056.559 & 5055.984       & Si~II (5)     & 0.1625 &    0.590 & 10.07$-$12.53 & 0.575 & $+$31.65 \\
5122.428 & 5121.69        & C~II (12)     & 0.0712 &          & 20.06$-$23.47 & 0.738 & $+$40.76 \\
5159.448 & 5158.81        & [Fe~II] (19F) & 0.049  &          &               & 0.638 & $+$34.63 \\
5194.567 & 5193.909       & Fe~III (5)    & 0.0186 & $-$2.852 & 8.66$-$11.04  & 0.658 & $+$35.54 \\
5198.512 & 5197.929       & Fe~I (1091)   & 0.0546 & $-$0.977 & 4.28$-$6.66   & 0.583 & $+$31.18 \\
5200.874 &                & UN            & 0.0217 &          &               &       &          \\ 
5244.009 & 5243.306       & Fe~III (113)  & 0.0512 &    0.405 & 18.27$-$20.63 & 0.703 & $+$37.75 \\
5262.238 & 5261.61        & [Fe~II] (19F) & 0.0375 &          &               & 0.628 & $+$33.34 \\
5273.955 & 5273.38        & [Fe~II] (18F) & 0.0182 &          &               & 0.575 & $+$30.24 \\
5282.946 & 5282.297       & Fe~III (113)  & 0.0350 &    0.108 & 18.27$-$20.61 & 0.649 & $+$34.39 \\
5299.612 & 5299.045       & O~I (26)      & 0.0501 & $-$2.140 & 10.99$-$13.33 & 0.567 & $+$29.63 \\
5343.07  &                & UN            & 0.0698 &          &               &       &          \\
5516.076 & 5515.335       & V~I (1)       & 0.0805 & $-$3.570 &  0.00$-$2.24  & 0.741 & $+$37.84 \\
5535.998 & 5535.346       & V~I (1)       & 0.0293 & $-$4.043 &  0.02$-$2.25  & 0.652 & $+$32.86 \\
5538.387 & 5537.760       & Mn~I (4)      & 0.0314 & $-$2.017 & 2.19$-$4.42   & 0.627 & $+$31.50 \\
5555.575 & 5554.94        & O~I (24)      & 0.0327 &          & 10.94$-$13.16 & 0.635 & $+$31.82 \\
5577.325 & 5576.61        & Si~II (9)     & 0.0332 &          &               & 0.715 & $+$35.99 \\
5577.832 & 5577.35        & [O~I] (3F)    & 0.0529 &          &               & 0.482 & $+$23.46 \\
5892.228 & 5891.598       & C~II (5)      & 0.1504 & $-$0.470 & 18.04$-$20.15 & 0.63  & $+$29.61 \\
5920.922 & 5920.124       & Fe~III (115)  & 0.0634 & $-$0.034 & 18.79$-$20.88 & 0.798 & $+$37.97 \\
5954.287 & 5953.613       & Fe~III (115)  & 0.0384 &    0.186 & 18.79$-$20.87 & 0.674 & $+$31.49 \\
5958.201 & 5957.559       & Si~II (4)     & 0.1694 & $-$0.300 & 10.07$-$12.15 & 0.642 & $+$29.86 \\
5959.277 & 5958.46        & O~I (23)      & 0.0285 &          & 10.94$-$13.01 &       &          \\
         &$+$5958.63      & O~I (23)      &        &          & 10.94$-$13.01 &       &          \\
5979.620 & 5978.930       & Si~II (4)     & 0.3998 &    0.000 & 10.07$-$12.15 & 0.69  & $+$32.15 \\
6000.211 & 5999.543       & Fe~III (117)  & 0.0563 &    0.355 & 18.82$-$20.88 & 0.668 & $+$30.93 \\
6033.247 & 6032.604       & Fe~III (117)  & 0.0830 &    0.497 & 18.82$-$20.87 & 0.643 & $+$29.51 \\
6047.098 & 6046.46        & O~I (22)      & 0.0835 &          & 10.94$-$12.98 & 0.638 & $+$29.18 \\
6095.944 & 6095.37        & C~II (24)     & 0.0490 &          & 22.47$-$24.50 & 0.574 & $+$25.78 \\
6099.236 & 6098.62        & C~II (24)     & 0.0837 &          & 22.47$-$24.50 & 0.616 & $+$27.83 \\
6152.043 &                & UN            & 0.1118 &          &               &       &          \\
6244.028 & 6243.355       & Al~II (10)    & 0.0428 &    0.670 & 13.08$-$15.06 & 0.673 & $+$29.87 \\
6257.864 &                & UN            & 0.0164 &          &               &       &          \\
6260.221 &                & UN            & 0.0242 &          &               &       &          \\
6300.877 & 6300.304       & [O~I] (1F)    & 0.6924 &          &               & 0.573 & $+$24.81 \\
6347.803 & 6347.109       & Si~II (2)     & 0.3636 &    0.300 & 8.12$-$10.07  & 0.694 & $+$30.33 \\
6364.348 & 6363.776       & [O~I] (1F)    & 0.2262 &          &               & 0.572 & $+$24.49 \\
6372.111 & 6371.371       & Si~II (2)     & 0.1646 &    0.000 &  8.12$-$10.07 & 0.74  & $+$32.37 \\   
6462.585 &                & UN            & 0.1393 &          &               &       &          \\ 

\hline
\end{tabular}
\end{center}
\end{table}

\setcounter{table}{2}
\begin{table}
\begin{center}
\renewcommand{\thetable}{\arabic{table}b}
\caption{contd...}
\begin{tabular}{|c|c|c|c|c|c|c|c|}
\hline
${\rm \lambda}_{\rm obs.}$ & ${\rm \lambda}_{\rm lab.}$ & Ident. & W$_{\rm \lambda}$ &
log (gf) & $\chi$ & $\Delta {\rm \lambda}$ & V$_{\rm r}$ \\
   (\AA~)            &   (\AA~)             &        &   (\AA~)       &
         &  (eV)  &      (\AA~)      & km s$^{-1}$  \\
\hline \hline
6546.69  & 6545.80        & Mg~II (23)    & 0.0218 &          & 11.58$-$13.47 & 0.89  & $+$38.32 \\
6548.82  & 6548.1         & [N~II] (1F)   & 0.1552 &          &               & 0.72  & $+$30.52 \\
         & 6583.6$^{\dagger}$ & [N~II](1F) &        &          &               &       &          \\ 
6611.462 & 6610.562       & N~II (31)     & 0.0364 &    0.430 & 21.60$-$23.48 & 0.9   & $+$38.37 \\
6667.584 & 6666.938       & O~II (85)     & 0.0253 & $-$1.030 & 28.94$-$30.80 & 0.646 & $+$26.60 \\   
6731.683 & 6730.79        & C~II (21)     & 0.0412 &          & 22.43$-$24.27 & 0.893 & $+$37.33 \\
6751.248 & 6750.22        & C~II (21)     & 0.0191 &          & 22.44$-$24.27 & 1.028 & $+$43.22 \\
6794.664 &                & UN            & 0.0641 &          &               &       &          \\   
6852.518 &                & UN            & 0.0321 &          &               &       &          \\
7002.964 & 7001.93        & O~I (21)      & 0.0958 &          & 10.94$-$12.70 &       &          \\
         & $+$7002.22     & O~I (21)      &        &          & 10.94$-$12.70 &       &          \\
7042.823 & 7042.048       & Al~II (3)     & 0.0695 &    0.350 & 11.32$-$13.08 & 0.775 & $+$30.55 \\
7053.906 & 7052.9         & C~II (26)     & 0.0323 &          & 22.80$-$24.55 & 1.006 & $+$40.32 \\
7057.413 & 7056.612       & Al~II (3)     & 0.0512 &    0.130 & 11.32$-$13.07 & 0.801 & $+$31.58 \\
7156.014 & 7155.14        & [Fe~II] (14F) & 0.0320 &          &               & 0.874 & $+$34.17 \\
7232.065 & 7231.332       & C~II (3)      & 0.801  &    0.070 & 16.33$-$18.04 & 0.733 & $+$27.94 \\
7237.489 & 7236.421       & C~II (3)      & 1.072  &    0.330 & 16.33$-$18.04 &       &          \\
         & $+$7236.91     & S~II (18)     &        &          & 14.09$-$15.80 &       &          \\
7255.214 & 7254.448       & O~I (20)      & 0.119  & $-$1.100 & 10.99$-$12.70 & 0.766 & $+$29.21 \\
7378.794 &                & UN            & 0.1842 &          &               &       &          \\ 
7412.59  &                & atmos.        & 0.0637 &          &               &       &          \\     
7443.104 & 7442.298       & N~I (3)       & 0.0369 & $-$0.450 & 10.33$-$11.99 & 0.806 & $+$30.02 \\
7469.175 & 7468.312       & N~I (3)       & 0.0555 & $-$0.270 & 10.33$-$11.99 & 0.863 & $+$32.20 \\
7750.613 &                & atmos.        & 0.0195 &          &               &       &          \\
7877.897 & 7877.054       & Mg~II (8)     & 0.1891 &    0.390 & 10.00$-$11.57 & 0.843 & $+$29.64 \\
7897.18  & 7896.367       & Mg~II (8)     & 0.3615 &    0.650 & 10.00$-$11.57 & 0.813 & $+$28.42 \\
8000.947 & 8000.12        & [Cr~II] (1F)  & 0.0765 &          &               & 0.827 & $+$28.54 \\
8126.292 & 8125.50        & [Cr~II] (1F)  & 0.0501 &          &               & 0.792 & $+$26.77 \\
8188.652 & 8187.95$^{*}$  & N~I (2)       & 0.0384 &          & 10.28$-$11.79 &       &          \\
8214.932 &                & UN            & 0.0631 &          &               &       &          \\
8217.301 & 8216.28        & N~I (2)       & 0.0582 &          & 10.29$-$11.79 & 1.021 & $+$34.81 \\
8224.161 & 8223.07        & N~I (2)       & 0.0979 &          & 10.29$-$11.79 & 1.091 & $+$37.33 \\

\hline
\end{tabular}

\noindent \parbox{14cm}{$^{\dagger}$ : [N~II](1F) 6583.6 \AA~ is blended with the
emission component of C~II(2) 6582.882 \AA~ P-Cygni profile. $^{*}$ : N~I (2) 8187.95 
emission line is affected by atmospheric absorption lines in the region.}

\end{center}
\end{table}

\setcounter{table}{2}
\begin{table}
\begin{center}
\renewcommand{\thetable}{\arabic{table}c}
\caption{Lines with P-Cygni profiles in IRAS17311-4924 (Hen3-1428). Equivalent
widths of the absorption and emission components of the P-Cygni profiles are
given. Wind velocities are estimated from the blue absorption edges of the
P-Cygni profiles.}
\begin{tabular}{|c|c|c|c|c|c|c|}
\hline
${\rm \lambda}_{\rm lab}$ & Ident. & W$_{\rm \lambda}$ (absorption) & W$_{\rm \lambda}$ (emission) & log (gf) &
$\chi$ & Wind Velocity\\
       (\AA~)       &        &  (\AA~)                    & (\AA~)                   &          &
(eV) & km s$^{-1}$\\
\hline \hline
5015.675             & He~I (4)     & 0.183 & 0.319 & $-$0.820 & 20.62$-$23.09 & $-$146\\
5073.78              & Fe~III (5)   & 0.015 & 0.009 & $-$2.557 & 8.65$-$11.09  & $-$81.36\\
5086.69              & Fe~III (5)   & 0.010 & 0.018 & $-$2.590 & 8.66$-$11.09  & \\
5127.463             & Fe~III (5)   & 0.070 & 0.025 & $-$2.218 & 8.65$-$11.07  & $-$81.74\\
5156.00              & Fe~III (5)   & 0.037 & 0.035 & $-$2.018 & 8.64$-$11.04  & $-$78.20\\
5875.618             & He~I (11)    &       & 0.992 &    0.410 & 20.97$-$23.08 &\\
$+$5875.650          & He~I (11)    &       &       & $-$0.140 & 20.97$-$23.08 &\\
$+$5875.989          & He~I (11)    &       &       & $-$0.210 & 20.97$-$23.08 &\\
6562.817             & H$_{\alpha}$ &       & 9.872 &    0.710 & 10.15$-$12.04 &\\
6578.03              & C~II (2)     & 0.327 & 0.082 &    0.120 & 14.45$-$16.33 & $-$114.20\\
6582.85$^{\dagger}$  & C~II (2)     & 0.269 &       & $-$0.180 & 14.45$-$16.33 &\\
6678.149             & He~I (46)    & 0.417 & 0.412 &    0.330 & 21.22$-$23.08 & $-$145.77\\
7065.188             & He~I (10)    & 0.137 & 1.062 & $-$0.460 & 20.96$-$22.72 &\\
$+$7065.719          & He~I (10)    &       &       & $-$1.160 & 20.96$-$22.72 &\\
7281.349             & He~I (45)    & 0.151 & 0.229 & $-$0.840 & 21.22$-$22.92 &\\

\hline
\end{tabular}

\noindent \parbox{14cm}{Two blended absorption components are visible
in the P-Cygni profile of He~I(11). The emission peak of the H$_{\alpha}$
P-Cygni profile is asymmetric indicating two emission components blended
together. Owing to the broad wings of the emission component in H$_{\alpha}$,
the absorption component of the P-Cygni profile lies above the normalised
continuum (see Fig. 3). $^{\dagger}$ : the emission component of C~II(2) 6582.85 \AA~
P-Cygni profile is blended with [N~II](1F) 6583.6 \AA~}

\end{center}
\end{table}

\setcounter{table}{2}
\begin{table}
\begin{center}
\renewcommand{\thetable}{\arabic{table}d}
\caption{Absorption (a) and emission (e) components of Na~I D$_{2}$ (5889.953 \AA~)
and Na~I D$_{1}$ (5895.923 \AA~) lines in the spectrum of IRAS17311-4924 (Hen3-1428).
W$_{\rm \lambda}$ are the equivalent widths of the components and V$_{\rm r}$ are the
respecitve heliocentric radial velocities.}
\begin{tabular}{|c|c|c|c|c|c|c|c|}
\hline
           & & \multicolumn{6}{c|}{IRAS17311-4924} \\ \cline{3-8}
          & & \multicolumn{3}{c|}{Na~I D$_{2}$} & \multicolumn{3}{c|}{Na~I D$_{1}$} \\ \cline{3-8}
Component & & ${\rm \lambda}_{\rm obs.}$ & W$_{\rm \lambda}$ & V$_{\rm r}$ &
${\rm \lambda}_{\rm lab}$ & W$_{\rm \lambda}$ & V$_{\rm r}$ \\
          & & (\AA~) & (\AA~) & (km s$^{-1}$) & (\AA~) & (\AA~) & (km s$^{-1}$) \\
\hline
1. & a & 5889.886 & 0.552 & $-$5.87  & 5895.875 & 0.479 & $-$4.90 \\
2. & a & 5890.334 & 0.121 & $+$16.94 & 5896.331 & 0.121 & $+$18.30 \\
3. & e & 5890.677 & 0.096 & $+$34.41 & 5896.695 & 0.020 & $+$36.82 \\

\hline
\end{tabular}
\end{center}
\end{table}

\begin{table}
\begin{center}
\caption{Expansion velocities}
\begin{tabular}{|c|c|c|c|c|c|c|c|}
\hline
\multicolumn{4}{|c|}{IRAS13266-5551} & \multicolumn{4}{c|}{IRAS17311-4924}\\
Ident. & ${\rm \lambda}_{\rm lab}$ & FWHM & V$_{\rm exp}$ & Ident. & 
${\rm \lambda}_{\rm lab}$ & FWHM & V$_{\rm exp}$ \\
       &                     & \AA~ & km s$^{-1}$ &     &     
                    & \AA~ & km s$^{-1}$ \\ 
\hline \hline
[N~II](1F)  &  6548.1          & 0.845 &  19.36      & [O~I](3F) & 5577.35 &
0.418 & 11.24 \\            
           &                  &     &               & [O~I](1F) & 6300.304 &
0.527 & 12.55 \\        
           &                  &     &               & [O~I](1F) & 6363.776 &
0.517 & 12.19 \\
           &                  &     &               & [N~II](1F) & 6548.1 & 
1.024 & 23.46 \\
\hline
\end{tabular}

\noindent \parbox{14cm}{From CO observations Loup et al. (1990) 
and Nyman et al. (1992) estimated expansion velocities of 
11 km s$^{-1}$ and 14.1 km s$^{-1}$ respectively in IRAS17311-4924.
Absorption component 2 of the Na~I profile in IRAS17311-4924
has radial velocity (= 18.30 km s$^{-1}$) comparable with the
expansion velocity of the star.}

\end{center}
\end{table}

\setcounter{table}{4}
\begin{table}
\begin{center}
\renewcommand{\thetable}{\arabic{table}a}
\caption{Derived chemical composition of IRAS13266-5551).
The abundances are for log $\epsilon$(H) = 12.0. Solar
abundances log $\epsilon$(X)$_{\odot}$ are from Grevesse \& Sauval (1998)
and Allende Prieto et al. (2001, 2002). 
n refers to the number of lines of each species used for the 
abundance determination. $\sigma$ is the standard deviation. 
For comparison we have listed the average 
abundances of main sequence B-stars from the Ori OB1 association (Kilian, 1992)}
\begin{tabular}{|c|c|c|c|c|c|c|c|c|c|c|}
\hline
 & \multicolumn{5}{c|}{IRAS13266-5551} & & \multicolumn{1}{c|}{Main sequence}\\
 & \multicolumn{5}{c|}{(23000, 3.0, 10)}             & & \multicolumn{1}{c|}{B-stars, Ori OB1}\\
X & n & log $\epsilon(X)$ & $\sigma$ & [X/H] & [X/Fe] & log $\epsilon$(X)$_{\odot}$ 
& log $\epsilon$(X) \\ \hline \hline

He~I   &  1 & 11.26$^{\dagger}$ &       & $+$0.33 & $+$0.50 & 10.93 & 11.04\\
C~II   &  4 &  8.67             & 0.27  & $+$0.28 & $+$0.45 &  8.39 &  8.23\\
N~II   &  6 &  8.23             & 0.29  & $+$0.31 & $+$0.48 &  7.92 &  7.72\\
O~II   &  4 &  8.78             & 0.28  & $+$0.09 & $+$0.26 &  8.69 &  8.60\\
Ne~I   &  4 &  9.16$^{*}$       & 0.32  & $+$1.08 & $+$1.25 &  8.08 &      \\
S~II   &  4 &  7.96$^{\dagger}$ &       & $+$0.63 & $+$0.80 &  7.33 &      \\
Fe~III &  4 &  7.33             & 0.32  & $-$0.17 &         &  7.50 &      \\

\hline
\end{tabular}

\vspace{0.2cm}

\noindent \parbox{16cm}{The abundances were derived using Kurucz's WIDTH9
program (modified for Unix machines by John Lester at the University of
Toronto, Canada). $\dagger$ : these values were derived from spectrum synthesis
analysis using the SYNSPEC code. $^{*}$: Ne abundances derived using Kurucz's
WIDTH9 program appear to be in error. The observed Ne~I lines are much stronger 
than the corresponding lines synthesised with the SYNSPEC code using the 
derived atmospheric parameters of the star and the Ne abundance in Table 5a. 
NLTE effects may be significant for these lines (see Sec. 3.8.5). }

\end{center}
\end{table}

\setcounter{table}{4}
\begin{table}
\begin{center}
\renewcommand{\thetable}{\arabic{table}b}
\caption{Uncertainities in the abundances, $\Delta$log $\epsilon$(X) 
due to uncertainities in the the model atmospheric parameters}
\begin{tabular}{|c|c|c|c|c|c|c|c|c|c|c|}
\hline
Element & $\Delta$T$_{\rm eff}$ & $\Delta$log $g$ & $\Delta\xi_{\rm t}$ & $\sigma_{\rm m}$ \\
        & $+$1000~K              & +0.5          & $+$ 1 km s$^{-1}$     &              \\
\hline \hline

C       & $+$ 0.12 &  $-$ 0.18 & $-$ 0.02 & 0.22 \\
N       & $+$ 0.12 &  $-$ 0.04 & $-$ 0.02 & 0.13 \\
O       & $-$ 0.03 &  $+$ 0.21 & $-$ 0.02 & 0.21 \\
Ne      & $+$ 0.16 &  $-$ 0.20 & $-$ 0.02 & 0.26 \\
Fe      & $+$ 0.14 &  $+$ 0.05 & $-$ 0.02 & 0.15 \\

\hline
\end{tabular}
\end{center}
\end{table}

\clearpage

\appendix

\setcounter{figure}{0}
\begin{figure}
\section{High resolution optical spectrum of IRAS 13266-5551 (CPD-55 5588)}
\renewcommand{\thefigure}{\Alph{figure}}
\epsfig{figure=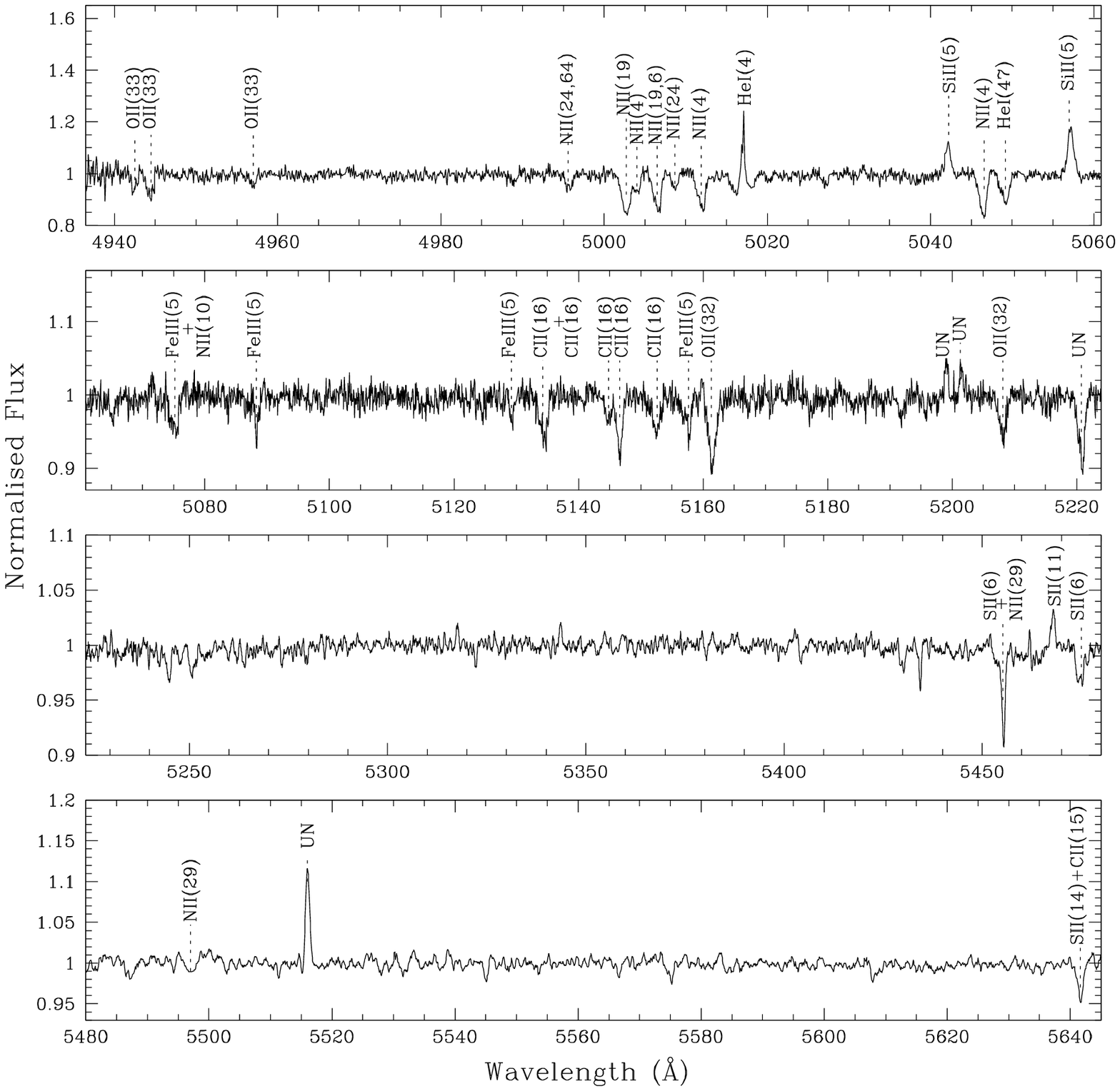, width=18cm, height=23cm}
\caption{Optical spectrum of IRAS13266-5551}
\end{figure}

\setcounter{figure}{0}
\begin{figure}
\renewcommand{\thefigure}{\Alph{figure}}
\epsfig{figure=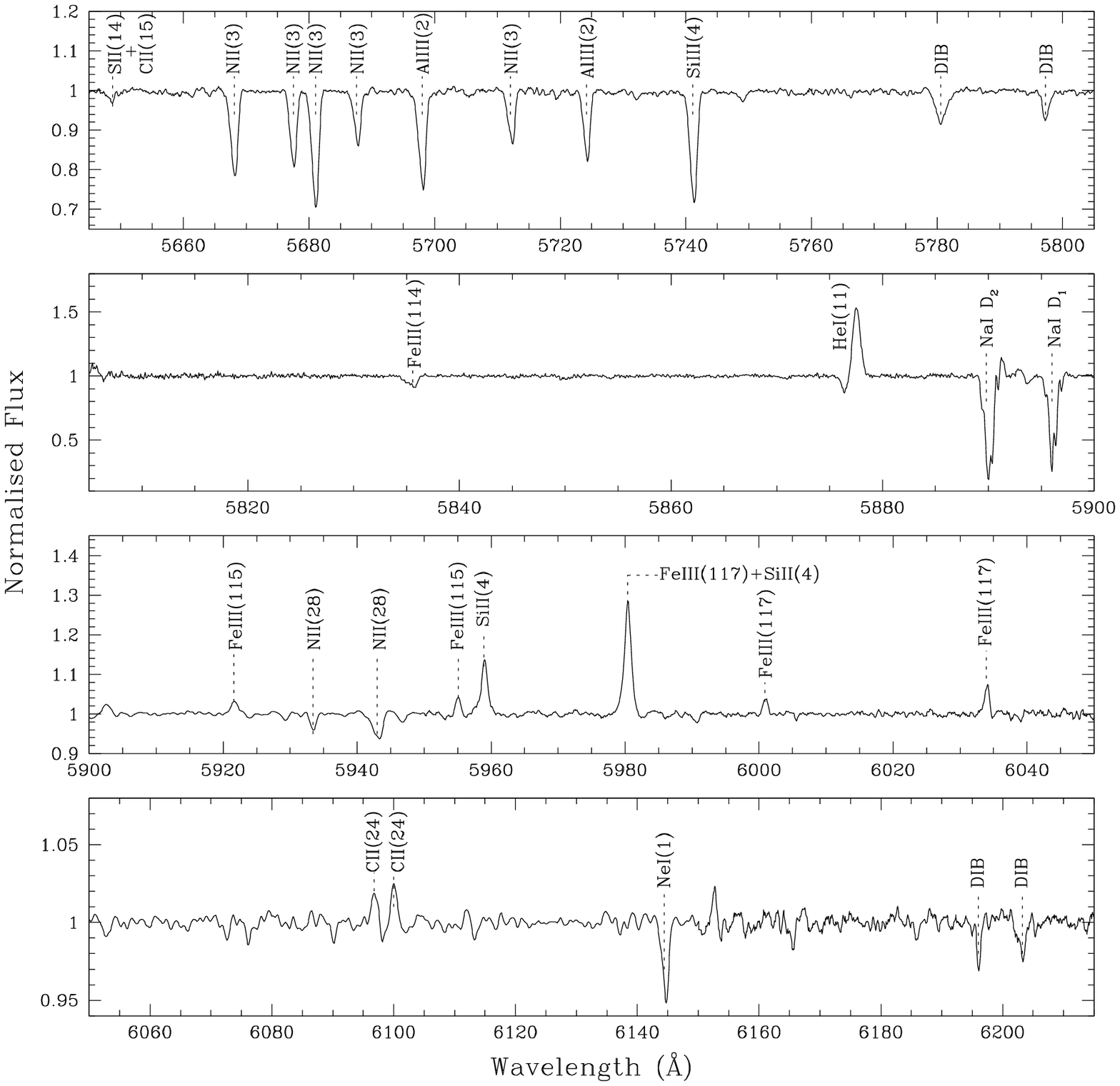, width=18cm, height=23cm}
\caption{Optical spectrum of IRAS13266-5551 contd...}
\end{figure}

\setcounter{figure}{0}
\begin{figure}
\renewcommand{\thefigure}{\Alph{figure}}
\epsfig{figure=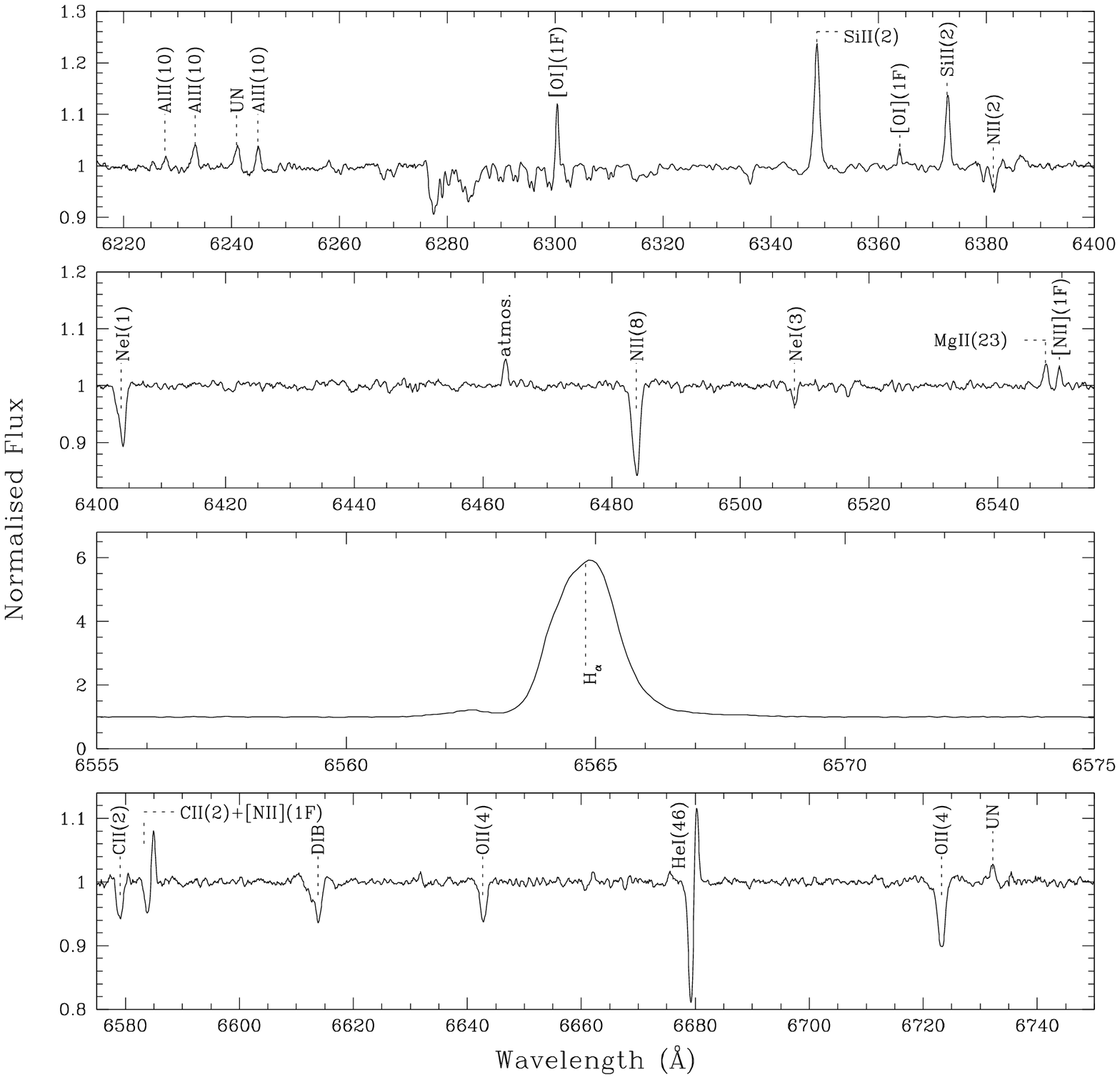, width=18cm, height=23cm}
\caption{Optical spectrum of IRAS13266-5551 contd...}
\end{figure}

\setcounter{figure}{0}
\begin{figure}
\renewcommand{\thefigure}{\Alph{figure}}
\epsfig{figure=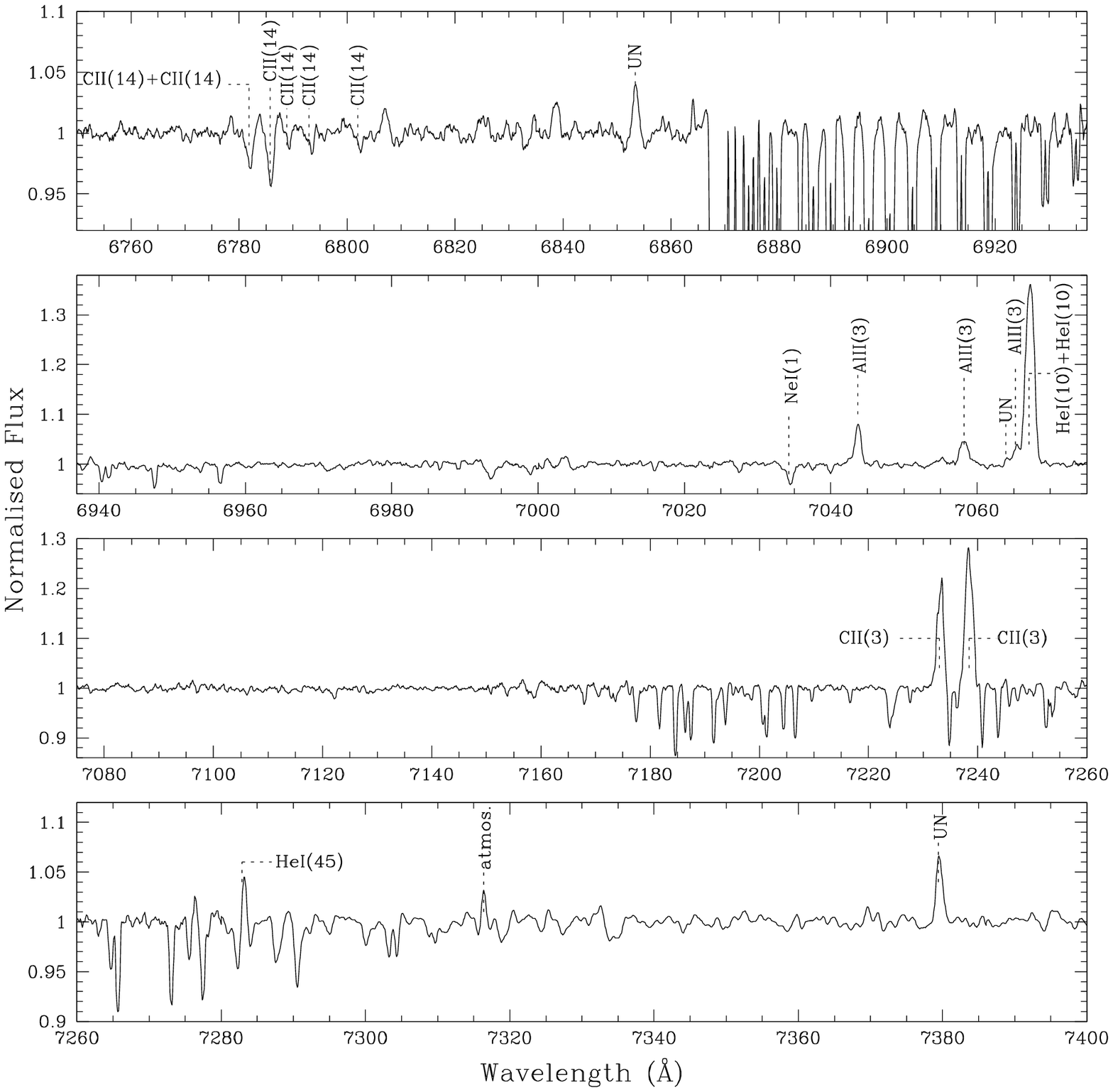, width=18cm, height=23cm}
\caption{Optical spectrum of IRAS13266-5551 contd...}
\end{figure}

\setcounter{figure}{0}
\begin{figure}
\renewcommand{\thefigure}{\Alph{figure}}
\epsfig{figure=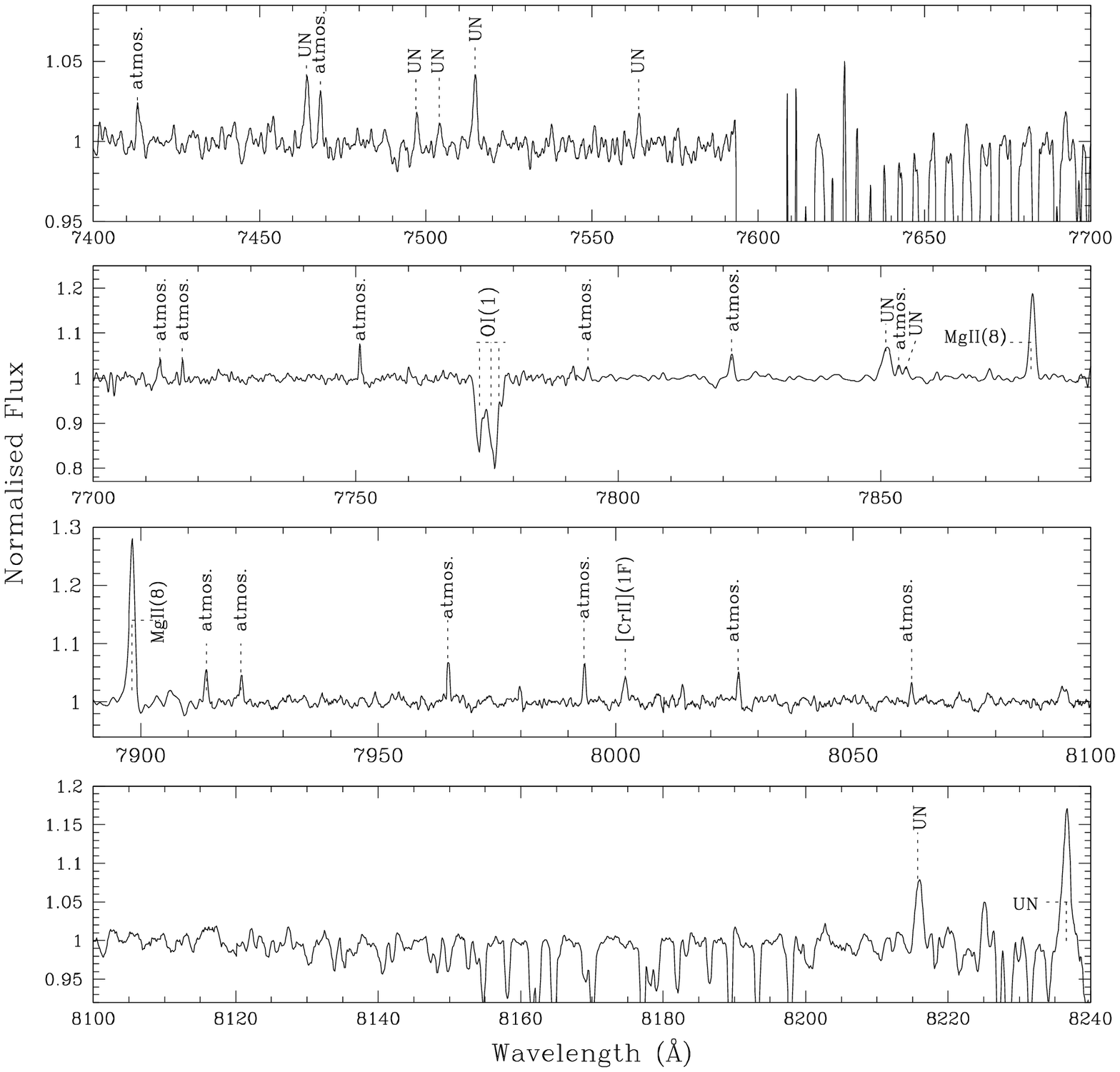, width=18cm, height=23cm}
\caption{Optical spectrum of IRAS13266-5551 contd...}
\end{figure}

\clearpage

\begin{figure}
\section{High resolution optical spectrum of IRAS17311-4924 (Hen3-1428)}
\setcounter{figure}{1}
\renewcommand{\thefigure}{\Alph{figure}}
\epsfig{figure=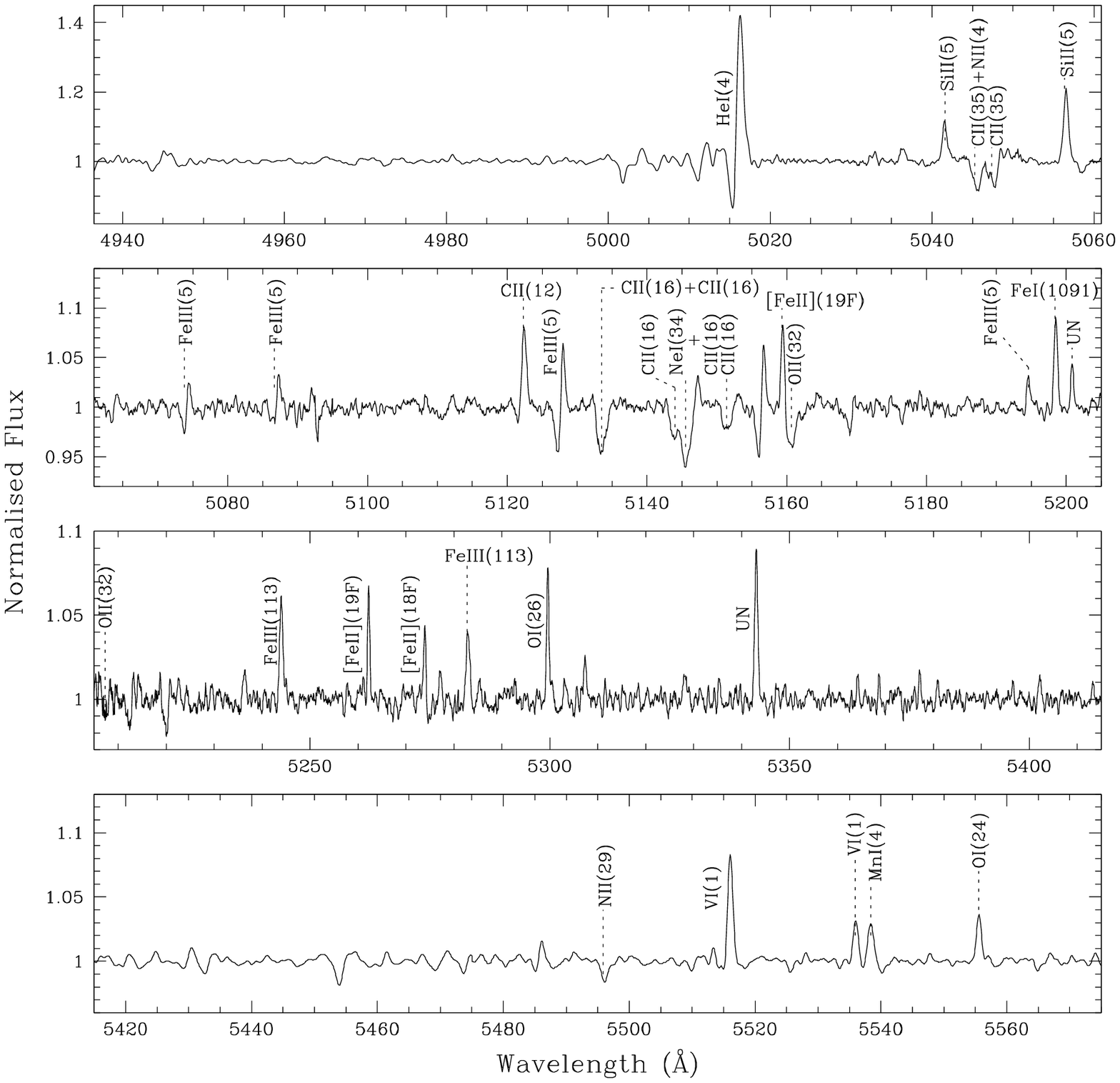, width=18cm, height=23cm}
\caption{Optical spectrum of IRAS17311-4924}
\end{figure}

\setcounter{figure}{1}
\begin{figure}
\renewcommand{\thefigure}{\Alph{figure}}
\epsfig{figure=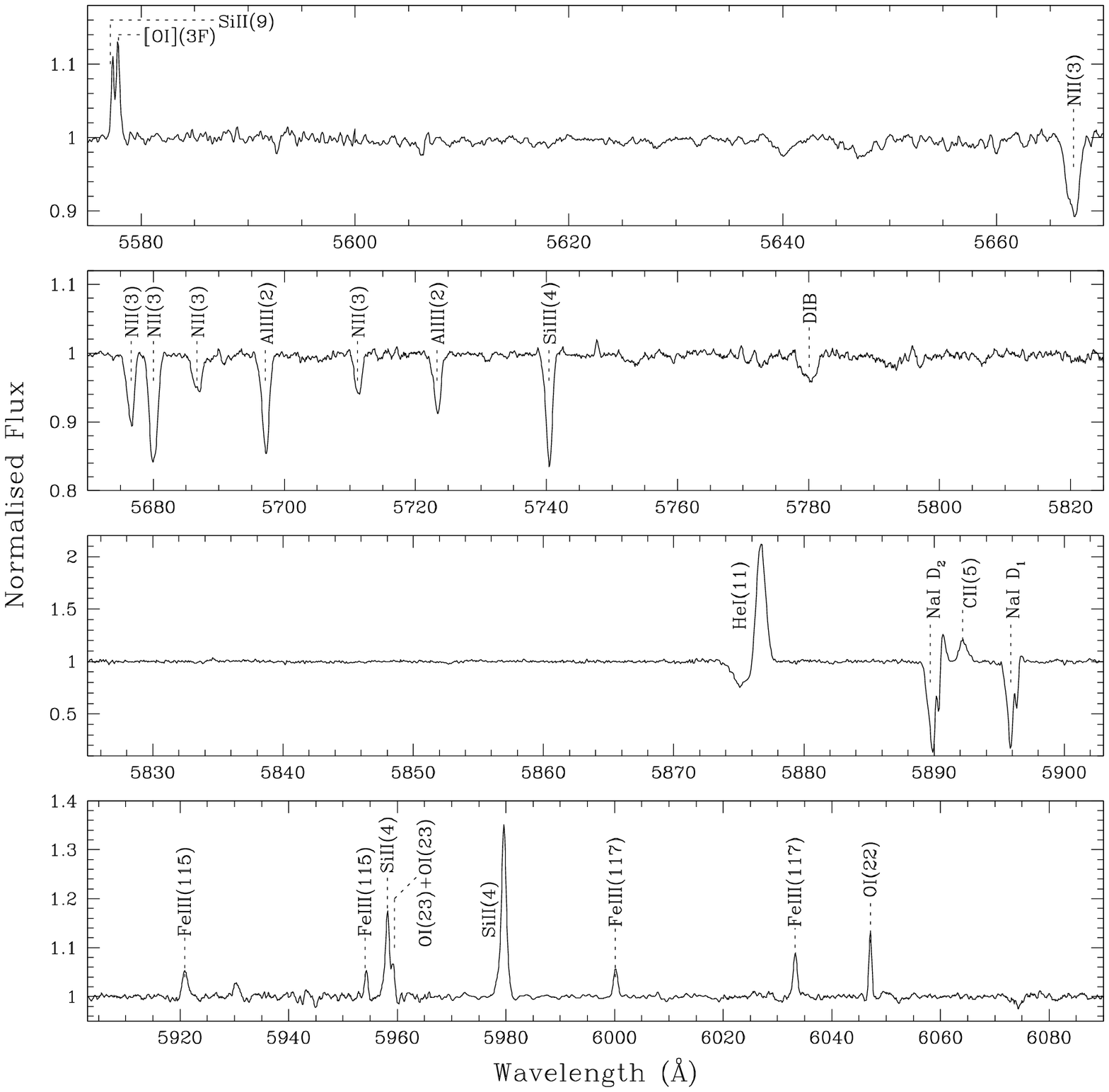, width=18cm, height=23cm}
\caption{Optical spectrum of IRAS17311-4924 contd...}
\end{figure}

\setcounter{figure}{1}
\begin{figure}
\renewcommand{\thefigure}{\Alph{figure}}
\epsfig{figure=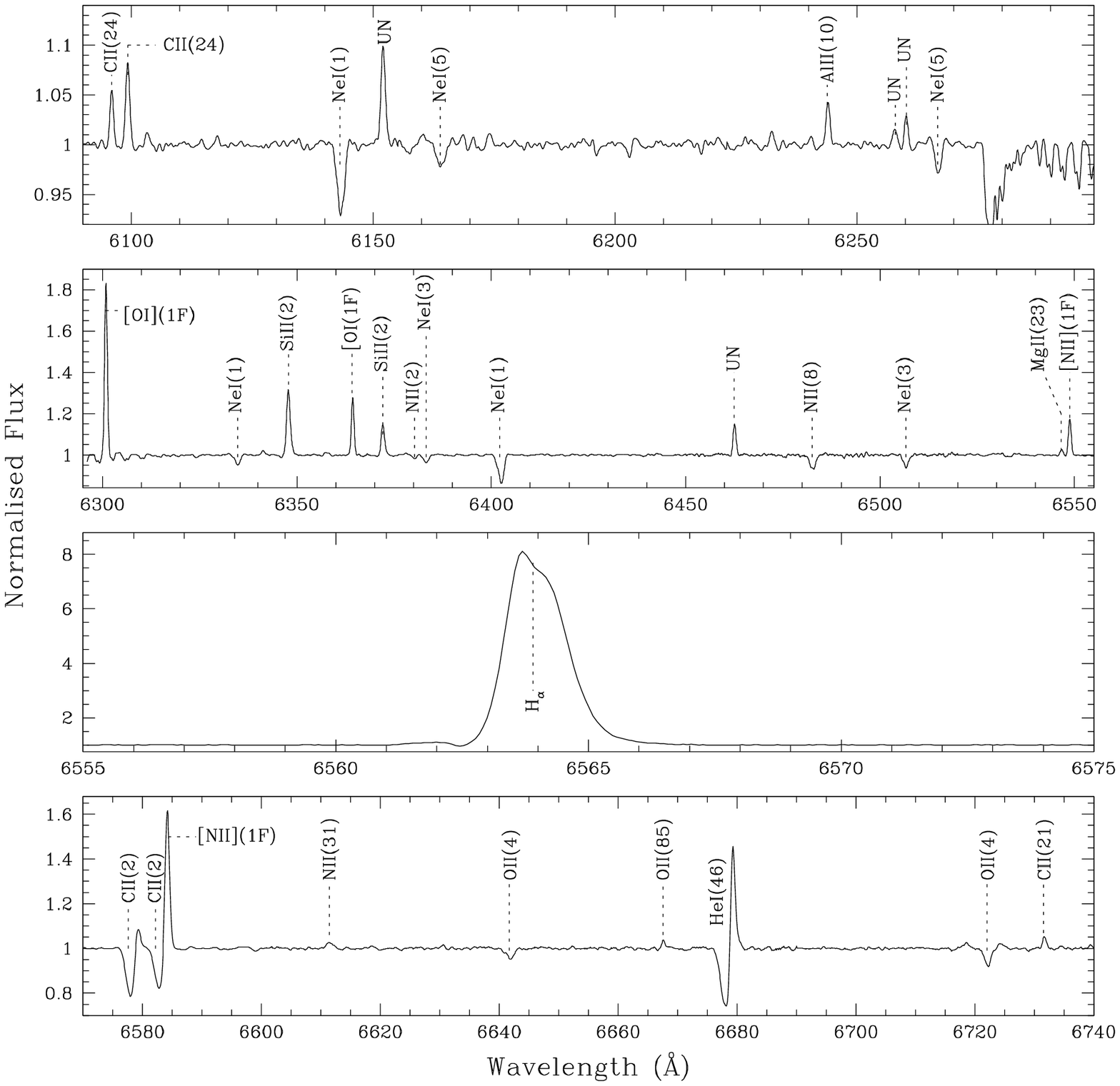, width=18cm, height=23cm}
\caption{Optical spectrum of IRAS17311-4924 contd...}
\end{figure}

\setcounter{figure}{1}
\begin{figure}
\renewcommand{\thefigure}{\Alph{figure}}
\epsfig{figure=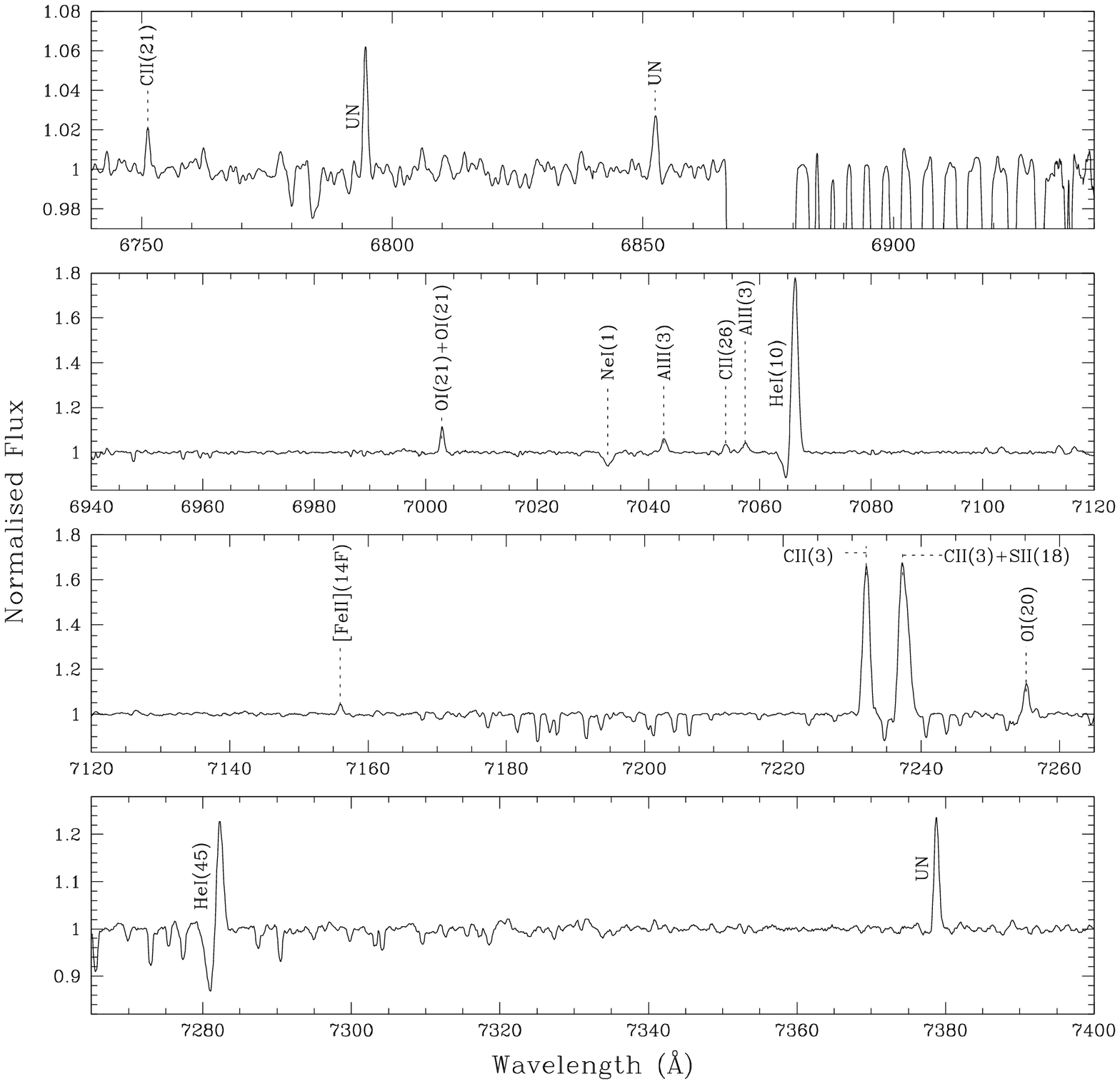, width=18cm, height=23cm}
\caption{Optical spectrum of IRAS17311-4924 contd...}
\end{figure}

\setcounter{figure}{1}
\begin{figure}
\renewcommand{\thefigure}{\Alph{figure}}
\epsfig{figure=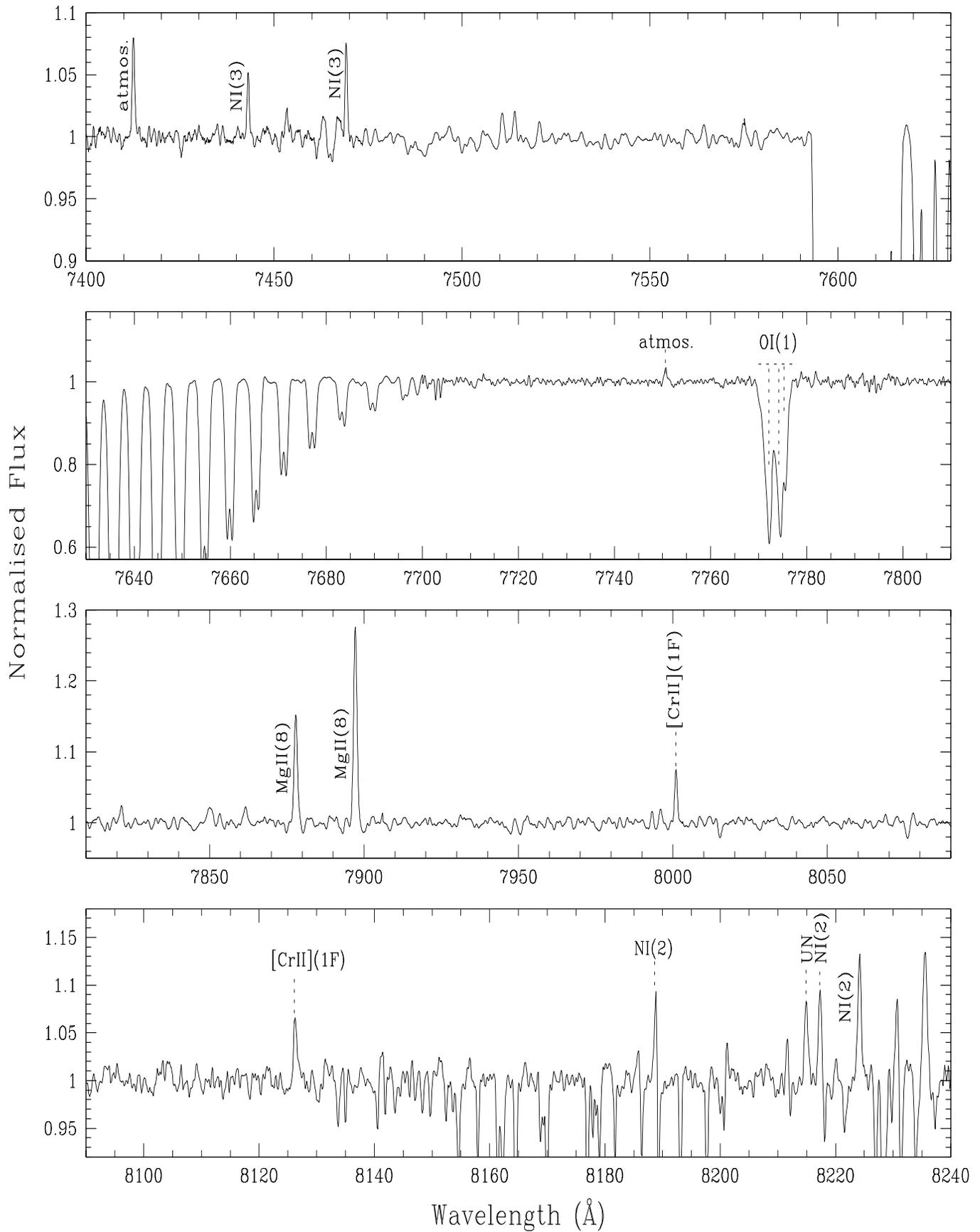, width=18cm, height=23cm}
\caption{Optical spectrum of IRAS17311-4924 contd...}
\end{figure}

\clearpage

\acknowledgements

We would like to thank Dr. John Lester at the University of Toronto,
Canada for having kindly provided the Unix version of the
WIDTH9 program. We would also like to thank the referee,
Dr. V. G. Klochkova for helpful comments.

\end{document}